\documentclass[aps,prd,twocolumn,showpacs,floatfix,
  superscriptaddress,amsmath]{revtex4}
\usepackage{epsf}              %
\usepackage{latexsym}           %
\usepackage{amscd}              %
\usepackage{amsmath}            
\usepackage{graphicx,subfigure} %
\usepackage{comment}            %
\input xy
\xyoption{all}

\renewcommand{\l}{\left}
\renewcommand{\r}{\right}

\newcommand{\beq}{\begin{equation}}
\newcommand{\eeq}{\end{equation}}
\newcommand{\bea}{\begin{eqnarray}}
\newcommand{\eea}{\end{eqnarray}}

\renewcommand{\o}{\overline}
\newcommand{\nn}{\nonumber}
\newcommand{\benn}{\begin{displaymath}}
\newcommand{\eenn}{\end{displaymath}}

\newcommand{\D}{{\Delta}}

\newcommand{\tw}{\textwidth}
\newcommand{\ig}{\includegraphics}
\def\slashchar#1{\ensuremath{                   %
   \setbox0=\hbox{${}#1{}$}                     
   \dimen0=\wd0                                 
   \setbox1=\hbox{/} \dimen1=\wd1               
   \ifdim\dimen0>\dimen1                        
   \rlap{\hbox to \dimen0{\hfil/\hfil}}         
   {}#1{}                                       
   \else                                        
   \rlap{\hbox to \dimen1{\hfil${}#1{}$\hfil}}  
   /                                            
   \fi}}                                        %
\def\simge{
 \mathrel{\rlap{\raise 0.511ex
  \hbox{$>$}}{\lower 0.511ex \hbox{$\sim$}}}}
\def\simle{
 \mathrel{\rlap{\raise 0.511ex
  \hbox{$<$}}{\lower 0.511ex \hbox{$\sim$}}}}

\begin{document}
 \include{def}
\preprint{DOE/ER/40762-398,JLAB-THY-07-707}

\title{Lattice QCD determination of patterns of excited baryon states}

\author{Subhasish Basak}
\affiliation{Department of Physics, Indiana University,
Bloomington, IN 47405}
\author{R.~G.~Edwards}
\affiliation{Thomas Jefferson National Accelerator Facility,
Newport News, VA 23606, USA}
\author{G.~T.~Fleming}
\affiliation{Yale University, New Haven, CT 06520, USA}
\author{K.~J.~Juge}
\affiliation{Department of Physics, University of the Pacific,
3601 Pacific Avenue Stockton, California 95211, USA}
\author{A.~Lichtl}
\affiliation{RIKEN BNL Research Center Building 510A, Physics
Department Brookhaven National Laboratory Upton, NY 11973-5000,
USA}
\author{C.~Morningstar}
\affiliation{Department of Physics, Carnegie Mellon University,
Pittsburgh, PA 15213, USA}
\author{D.~G.~Richards}
\affiliation{Thomas Jefferson National Accelerator Facility,
Newport News, VA 23606, USA}
\author{I.~Sato}
\affiliation{Lawrence Berkeley Laboratory, Berkeley, CA 94720,
USA}
\author{S.~J.~Wallace}
\affiliation{University of Maryland, College Park, MD 20742, USA}

\begin{abstract}
Energies for excited isospin $I=\frac{1}{2}$ and
$I=\frac{3}{2}$ states that include the nucleon and $\Delta$
families of baryons are computed using quenched, anisotropic
lattices. Baryon interpolating field operators that are used
include nonlocal operators that provide $G_2$ irreducible
representations of the octahedral group.  The decomposition of
spin $\frac{5}{2}$ or higher spin states is realized for the first
time in a lattice QCD calculation. We observe patterns of
degenerate energies in the irreducible representations of the
octahedral group that correspond to the subduction of the
continuum spin $\frac{5}{2}$ or higher.  The overall pattern of
low-lying excited states corresponds well to the pattern of
physical states subduced to the irreducible representations of the
octahedral group.
\end{abstract}
 \maketitle
\section{Introduction}
The theoretical determination of the spectrum of baryon resonances
from the fundamental quark and gluon degrees of freedom is an
important goal for lattice QCD. To date there have been many
lattice studies of ground state energies for different
baryons~\cite{Aoki:1999yr,Davies:2003ik,AliKhan:2001tx} but only a
few results for excited state energies have been
reported.~\cite{Zhou:2006xe,Burch:2006cc,Sasaki:2005uq,
Sasaki:2005ug,Basak:2004hr,Zanotti:2003fx,Guadagnoli:2004,Lichtl-Thesis} No clear
determination of states with spin $\frac{5}{2}$ or higher has been
published because nonlocal operators have not been used. In this
work we find degenerate energies that occur in irreducible
representations of the octahedral group corresponding to the
subduction of the continuum spin $\frac{5}{2}$ or higher. The
pattern of lattice QCD baryon states that we observe is compared
with the physical spectrum.

Lattice correlation functions correspond to definite irreducible
representations (irreps) of the octahedral group when the source
and sink operators transform accordingly. There are six
double-valued irreps of the octahedral group: three for
even-parity that are labeled with a $g$ subscript ({\it gerade})
and three for odd-parity that are labeled with a $u$ subscript
({\it ungerade}). They are: $G_{1g}, H_g, G_{2g}, G_{1u}, H_u$ and
$G_{2u}$.

Continuum values of total angular momenta are realized in lattice
simulations by patterns of degenerate energies in the continuum
limit that match the patterns in Table~\ref{table:subduction} for
the subduction of spin $J$ to the double-valued irreps of the
octahedral group.
\begin{table}
\caption{The number of occurrences of double-valued irrep
$\Lambda$ of the octahedral group for half-integer values of
continuum spin $J$. } \label{table:subduction}
\begin{tabular}{|cllllllll|}
\hline
$\Lambda$   &  &J= & ~$1/2$~&~$3/2$~&~ $5/2$ ~&~ $7/2$~&~ $9/2$~&~ $11/2$ \\
\hline
$G_{1}$ ~&~  ~&~  ~&~ 1 ~&~ 0 ~&~ 0 ~&~ 1 ~&~ 1 ~&~ 1     \\
$H $    ~&~  ~&~  ~&~ 0 ~&~ 1 ~&~ 1 ~&~ 1 ~&~ 2 ~&~ 2  \\
$G_{2}$ ~&~  ~&~  ~&~ 0 ~&~ 0 ~&~ 1 ~&~ 1 ~&~ 0 ~&~ 1   \\
\hline
\end{tabular}
\end{table}
For example, a state in one of the $G_2$ irreps is a signal for
the subduction of continuum spin $\frac{5}{2}$ or higher. For spin
$\frac{5}{2}$, there must be partner states in the $H$ and $G_2$
irreps that would be degenerate in the continuum limit. For spin
$\frac{7}{2}$, there must be partner states in the $G_1$, $H$ and
$G_2$ irreps.

This paper reports on work to determine the pattern of low-lying
states in the $I=\frac{1}{2}$ and $I=\frac{3}{2}$
channels. We carry out an analysis in quenched lattice QCD using
quasi-local and one-link-displaced operators formed from three
quark fields. Smeared quark and gluon fields are
used.~\cite{Albanese87,Alford:1995dm} Smearing reduces the
couplings to short wavelength fluctuations of the theory and
provides cleaner determinations of energies.\cite{Lichtl-Dublin}
This is important when a large array of interpolating field
operators is used in order to implement the variational method of
Refs.~\cite{michael85,lw90}.

Both quasi-local and nonlocal operators transforming according to
the $G_2$ irrep are required. We have developed sets of baryon
operators that transform according to irreducible representations
in two publications, one using an analytical method based on
appropriate Clebsch-Gordan coefficients for the octahedral
group~\cite{Basak:2005ir} and the other using a projection method
that has been automated and provides very large sets of
operators~\cite{Basak:2005aq}. Results obtained from both methods
are found to agree. In this work we use, for the positive parity channels, the
three-quark operators defined in Tables VI, VII, and X of
Ref.~\cite{Basak:2005ir}. These compose a complete set of
quasi-local operators plus the simplest set of nonlocal operators
having one quark displaced relative to the other two.

The level ordering of baryon masses has been modelled using the
spin-flavor SU(6) quark model.
Negative parity multiplets are degenerate under SU(6) symmetry but
the symmetry is broken by addition of a spin-spin contact
interaction between quarks~\cite{Capstick:2000qj}. As a result,
the lowest $N(J^P={1\over 2}^-)$ and $N({3\over 2}^-)$ states
become less massive than the unperturbed levels and other states,
such as $N({5\over 2}^-)$, $(\Delta{1\over 2}^-)$, and $\Delta
({3\over 2}^-)$, become more massive. This pattern is observed in
our lattice calculations. Degeneracies of SU(6) are further broken
by addition of a spin-spin tensor interaction. In the positive
parity excited states, Isgur and Karl~\cite{Isgur:1978wd}
introduced an anharmonic perturbation in order to explain the mass
splittings, and a hyperfine interaction was introduced to break the
degeneracy of the multiplets. They found that the second excited
$N({1\over 2}^+)$ state, the lowest $N({3\over 2}^+)$ state and
the $N({5\over 2}^+)$ states are nearly degenerate. Our lattice
results using a relatively large pion mass equal to
$490\;\mbox{MeV}$ follow this pattern.

This paper is organized as follows. In
section~\ref{sec:improvec_baryon_operators}, we review the baryon
operator construction and show the types of operators used in our
simulations. In section~\ref{sec:charge-conjugation} the use of
charge conjugation together with time reversal is discussed. The
combination provides a means to obtain a second correlation matrix
for each gauge configuration.  In
section~\ref{sec:computational_techniques} we review the
computational techniques with special emphasis on the variational
method that is essential for obtaining excited states. Also we
discuss the anisotropic action that is employed and the tuning of
parameters required to obtain the correct speed of light.
Section~\ref{sec:results} presents and discusses the results for
energies and eigenvectors in each irrep and their stability with
respect to the number of operators used in the variational method.
Section~\ref{sec:patterns} discusses the pattern of lowest energy
levels in the lattice results and shows that it is consistent with
the physical spectrum.  Two notable findings are: 1.)
the first positive-parity excited state in the $I=\frac{1}{2}, G_{1g}$ channel is
found to have energy significantly higher than the
experimentally-known mass of the Roper resonance, and 2.) spin
assignments of either $\frac{5}{2}$ or $\frac{7}{2}$ are possible
for some lattice levels that preferably would be assigned spin
$\frac{5}{2}$ in order to match the pattern of the physical
levels. Section~\ref{sec:summary} gives a summary of the results.

\section{Improved baryon operators}
\label{sec:improvec_baryon_operators}

\subsection{Quasi-local operators}
\label{subsec:quasi-local_operators}

We employ operators that transform as irreducible representations of
the octahedral group.  They are formed as linear combinations
of elemental operators that have definite isospin
and strangeness as follows,
\renewcommand{\a}{\alpha}
\renewcommand{\b}{\beta}
\renewcommand{\c}{\gamma}
\begin{equation}
\o{B}^{(\Lambda\lambda)}_k({\bf x},t) =
c^{(\Lambda\lambda k)}_{\a\b\c,0}
\o{B}_{\a\b\c}({\bf x},t),
\label{eq:irrep_op_construction}
\end{equation}
where $\a,\b,\c$ are Dirac indices and the Clebsch-Gordan
coefficients for the octahedral group provide the appropriate
coefficients $c_{\a\b\c,0}^{(\Lambda\lambda
k)}$.~\cite{Basak:2005ir} The subscript `0' denotes quasi-local
operators in which all quark fields have the same coordinates,
$({\bf x}, t)$. Octahedral group irreps are denoted by $\Lambda=\{
G_{1g}, G_{2g}, H_g, G_{1u}, G_{2u}, H_u\}$ and the corresponding
irrep dimensions are $d_\Lambda =\{ 2,2,4,2,2,4\}$, respectively.
Each baryon operator carries a row label that distinguishes
between the $d_\Lambda$ members of irrep $\Lambda$, i.e.,
$\lambda=1,2,\cdots,d_\Lambda$. When group representations contain
$m_\Lambda$ occurrences of irrep $\Lambda$, the label $k$ is used
to denote the $k$-th embedding.

Quasi-local elemental operators $\o{B}_{\a\b\c}$ in
Eq.~(\ref{eq:irrep_op_construction}) are color singlet
combinations of products of three quark fields, each of which is
smeared in the same gauge-covariant fashion about point ${\bf x}$.
They are listed in Table~\ref{table:quasi-local} for the different
baryons. Although not shown in the table, each quark field has a
color index $i$, $j$ or $k$, and a factor $\epsilon_{ijk}$ is
included in order to construct a color-singlet combination.
 The Dirac index symmetries that give non-vanishing
operators and the number of available quasi-local operators
distributed over gerade irreps are also shown in the table.

\begin{table}
\begin{center}
\caption{Baryons and the corresponding quasi-local three-quark
elemental operators. Columns 1 to 4 show the symbol, isospin,
strangeness, and the form of elemental quasi-local source
operators with maximum $I_z$, respectively. In the next column,
the MA label (mixed-antisymmetric) denotes combinations of three
Dirac indices that are antisymmetric under permutation of the
first two labels and orthogonal to the totally antisymmetric
combination. Similarly, the MS label (mixed-symmetric) denotes
combinations of indices that are symmetric with respect to the
first two labels and orthogonal to the totally symmetric
combination. Columns 6-8 show numbers of embeddings of operators
with irreps $G_{1g}, G_{2g}$ and $H_{g}$, respectively. The last
row shows the total number of even-parity operators, including all
rows. Odd-parity operators are obtained from the even-parity ones
as explained in the text.
  }
\begin{ruledtabular}
\begin{tabular}{c|cccccccc}
  $\o{B}$    & $I$ & $S$ & $\o{B}_{\alpha\beta\gamma}$&
Dirac  & $G_{1g}$ & $G_{2g}$ & $H_{g}$ & Total \\
\hline $\o{N}$ &      $1/2$ & $0$ & $(\o u_\a \o d_\b - \o d_\a \o
u_\b) \o u_\c/\sqrt{2}$ &
MA  &  3  &  0   &  1  & 10 \\
$\o{\Delta}$&  $3/2$ & $0$ & $\o u_\a \o u_\b \o u_\c$ &
S   &  1  &  0   &  2  & 10 \\
$\o{\Lambda}$& $0$   &$-1$ & $(\o u_\a \o d_\b - \o d_\a \o u_\b) \o
s_\c /\sqrt{2}$ &
MA,A&  4  &  0   & 1   & 12 \\
$\o{\Sigma}$  & $1$  &$-1$ & $\o u_\a \o u_\b \o s_\c$ &
MS,S&  4  &  0   & 3   & 20 \\
$\o{\Xi}$  &   $1/2$ &$-2$ & $\o s_\a \o s_\b \o u_\c$ &
MS,S&  4  &  0   & 3   & 20 \\
$\o{\Omega}$  & $0$  & $3$ & $\o s_\a \o s_\b \o s_\c$ &
S   &  1  &  0   & 2   & 10
\end{tabular}
\end{ruledtabular}
\label{table:quasi-local}
\end{center}
\end{table}

Note that the ``barred'' form of the operator is given
and it is formed from three ``barred'' quark fields. The
``unbarred'' form of baryon operator is obtained by a similar sum
using complex conjugates of the same coefficients and
three quark fields,
\begin{equation}
B^{(\Lambda\lambda)}_k({\bf x},t)= c^{(\Lambda\lambda k)*}_{\alpha
\beta \gamma,0} B_{\alpha \beta\gamma}({\bf x},t).
\end{equation}
Note that no $G_2$ irreps are obtained from quasi-local
interpolating fields.

A quark field operator and a ``barred'' field operator
are transformed into one another by charge conjugation
as follows,
\begin{align}
{\cal C}\o q_\a {\cal C}^\dag = -C_{\a\a'} q_{\a'},~~~
{\cal C}   q_\a {\cal C}^\dag = \o q_{\a'}C^\dag_{\a'\a},
\end{align}
where $C\equiv\gamma_4\gamma_2$.
A three-quark operator that transforms according to irrep $\Lambda$
of the octahedral group is related by charge conjugation to an
operator that transforms according to irrep $\Lambda_c$, which has opposite
parity, i.e.,
\begin{align}
& {\cal C} \o B_k^{(\Lambda\lambda)} {\cal C}^\dag
= c_{\a\b\c,0}^{(\Lambda\lambda k)} {\cal C} \o B_{\a\b\c} {\cal C}^\dag \nn\\
& = -c_{\a\b\c,0}^{(\Lambda\lambda k)} {\bf C}_{\a\b\c\a'\b'\c'} B_{\a'\b'\c'}
\equiv -c^{(\Lambda_c \lambda_c k)}_{\a'\b'\c',0} B_{\a'\b'\c'}, \nn \\
& {\cal C} B_k^{(\Lambda\lambda)} {\cal C}^\dag
= c_{\a\b\c,0}^{(\Lambda\lambda k)*} {\cal C} B_{\a\b\c} {\cal C}^\dag \nn \\
& = c_{\a\b\c,0}^{(\Lambda\lambda k)*} \o B_{\a'\b'\c'}
{\bf C}^\dag_{\a'\b'\c'\a\b\c}
= c^{(\Lambda_c \lambda_c k)*}_{\a'\b'\c',0} \o B_{\a'\b'\c'},
\label{eq:baryon_charge_conj}
\end{align}
where ${\bf C}_{\a'\b'\c'\a\b\c}=C_{\a'\a}C_{\b'\b}C_{\c'\c}$. If
$\Lambda$ is $G_{1g}, G_{2g}$ or $H_g$, then $\Lambda_c$ is
$G_{1u}, G_{2u}$ or $H_u$, respectively, and $\lambda_c=d_\Lambda
+1 - \lambda$. This one-to-one correspondence between an operator
of one parity and an operator with the opposite parity holds for
any quasi-local or displaced operator. Choosing the gerade
operators according to Eq.~(\ref{eq:irrep_op_construction}) and
the ungerade operators according to the right side of
Eq.~(\ref{eq:baryon_charge_conj}) provides a convenient labeling
of operators that facilitates the use of charge-conjugation
relations to realize improved statistics, as explained in
Sec.~\ref{sec:charge-conjugation}.

\subsection{One-link-displaced operators}
\label{subsec:one-link_operators}

A simple extension of quasi-local operators involves displacement
of the third quark field along a spatial direction relative to the
first two quark fields, with a gauge link included in order to
maintain gauge covariance. An abbreviated notation for
displacements uses the operator $\hat{d}_{\ell}$ whose action on a
three-quark quasi-local operator is defined as follows,
\begin{align}
\hat{d}_{\ell} \, &\o{B}^{(\Lambda\lambda)}_k({\bf x},t) =
\epsilon_{ijk} f^{(IS)}_{abc} c^{(\Lambda\lambda k)}_{\a\b\c}
\nn\\
& \times \o{\tilde{q}}_\a^{i, a}({\bf x},t) \o{\tilde{q}}_\b^{j,
b}({\bf x},t) \o{\tilde{q}}_\c^{k',c}({\bf x}+\hat{\ell},t)
U^{\dag k'k}_{\ell}({\bf x},t), \label{eq:displaced-ops}
\end{align}
where $\hat{\ell}$ can take any of six spatial directions, $\pm x, \pm
y, \pm z$.

Displacement operators transform amongst themselves under lattice
rotations. Linear combinations of displacements provide suitable
bases for the construction of irreps. We define six linearly
independent combinations of displacements and the operators that
create them as follows,
\begin{align}
&
\begin{pmatrix}
  \hat{A}_1 \o{B} &
  \hat{D}_+ \o{B} &
  \hat{D}_- \o{B} &
  \hat{D}_0 \o{B} &
  \hat{E}_0 \o{B} &
  \hat{E}_2 \o{B}
\end{pmatrix}
^T \equiv \nn\\ &
\begin{pmatrix}
  {1\over \sqrt{6}}( \hat{d}_x\o{B} + \hat{d}_y\o{B} +
  \hat{d}_z\o{B} + \hat{d}_{-x}\o{B} +
  \hat{d}_{-y}\o{B} + \hat{d}_{-z}\o{B}) \\
      {i \over 2}[ (\hat{d}_x \o{B} - \hat{d}_{-x}\o{B})
    + i( \hat{d}_y \o{B} - \hat{d}_{-y}\o{B})] \\
      -{i\over 2}[(\hat{d}_x\o{B} - \hat{d}_{-x}\o{B})
    - i(\hat{d}_y\o{B} - \hat{d}_{-y}\o{B})] \\
      -{i\over \sqrt{2}}(\hat{d}_z\o{B} - \hat{d}_{-z}\o{B})\\
    \!\! {1\over \sqrt{12}}[2(\hat{d}_z\o{B} + \hat{d}_{-z}\o{B})
      -(\hat{d}_x\o{B} + \hat{d}_{-x}\o{B}) -
      (\hat{d}_y\o{B} + \hat{d}_{-y}\o{B})] \!\!\! \\
    {1\over 2}[(\hat{d}_x\o{B} + \hat{d}_{-x}\o{B}) -
      (\hat{d}_y\o{B} + \hat{d}_{-y}\o{B})]
\end{pmatrix}
\label{eq:irrep-displacements}
\end{align}
where displacement operators denoted as $\hat{A}_1,
\hat{D}_{\pm,0}$, and $\hat{E}_{0,2}$ create combinations of
operators that transform according to the $A_1$, $T_1$, and $E$
single-valued irreps of the octahedral group, respectively.
\begin{table}
\caption{The number of occurrences of single-valued irrep
$\Lambda$ of the octahedral group for integer values of continuum
angular momentum $L$. } \label{table:subduction-2}
\begin{tabular}{|cclllll|}
\hline
$\Lambda$  &  L= & ~0~ &~1~  &~ 2 ~&~3~  &~ 4~     \\
\hline
$A_{1}$   ~&~   ~&~ 1 ~&~ 0 ~&~ 0 ~&~ 0 ~&~ 1 ~    \\
$T_1 $    ~&~   ~&~ 0 ~&~ 1 ~&~ 0 ~&~ 1 ~&~ 1 ~    \\
$E$       ~&~   ~&~ 0 ~&~ 0 ~&~ 1 ~&~ 0 ~&~ 1 ~    \\
$T_2 $    ~&~   ~&~ 0 ~&~ 0 ~&~ 1 ~&~ 1 ~&~ 1 ~    \\
$A_2$     ~&~   ~&~ 0 ~&~ 0 ~&~ 0 ~&~ 1 ~&~ 0 ~    \\
\hline
\end{tabular}
\end{table}
As may be seen in Table~\ref{table:subduction-2}, the lowest
angular momentum in $A_1$ is $L = 0$, the lowest in $T_1$ is $L=1$
and the lowest in $E$ is $L=2$. The combinations of displacements
are chosen so that they transform as the lattice discretizations
of the spherical harmonics $Y_{L m}$, {\it i.e.}, $\hat{A}_1 \sim
Y_{00}$, $\hat{D}_{+,0,-} \sim Y_{11}, Y_{10}, Y_{1-1}$, and
$\hat{E}_{0,2} \sim Y_{20}, (Y_{22}+Y_{2-2})$. Using this
convention, the $A_1$, $T_1$, and $E$ one-link-displaced operators
are defined as
\begin{align}
\o{B}^{(\Lambda\lambda)}_k =& \l\{
\begin{array}{l}
\hat{A}_1 \o{B}^{(\Lambda\lambda)}_{k'} \\
\sum_{r,\lambda'} C
\renewcommand{\arraystretch}{0.6}
\begin{pmatrix}\Lambda & T_1 & \Lambda' \\ \lambda & r & \lambda'\end{pmatrix}
\hat{D}_r \o{B}^{(\Lambda'\lambda')}_{k'}, \\
\sum_{r,\lambda'} C
\renewcommand{\arraystretch}{0.6}
\begin{pmatrix}\Lambda & E & \Lambda' \\ \lambda & r & \lambda'\end{pmatrix}
\hat{E}_r \o{B}^{(\Lambda'\lambda')}_{k'}
\end{array}
\r. \nn\\
\equiv & c^{(\Lambda\lambda k)}_{\a\b\c, \ell} \hat{d}_{\ell} \o
B_{\a\b\c} \label{eq:one-link-irreps}
\end{align}
respectively, where
$C
\renewcommand{\arraystretch}{0.6}
\begin{pmatrix}
\Lambda & \Lambda' & \Lambda'' \\
\lambda & \lambda' & \lambda''
\end{pmatrix}$
are Clebsch-Gordan coefficients appropriate for forming overall
irrep $\Lambda$ from direct products of irreps $\Lambda'$ and
$\Lambda''$.  The coefficients follow the conventions of
Ref.~\cite{Basak:2005ir}.
\begin{table}[h]
\begin{center}
\caption{Numbers of embeddings of even-parity irreps,  $G_{1g}$,
$G_{2g}$, and $H_g$, that are obtained with one-link-displaced
operators for different baryons.  Columns show the number of
operators for each overall irrep that can be made using the three
irreps of one-link displacements: $A_1$, $T_1$, and $E$. Operators
corresponding to a total derivative acting on a three-quark
operator are omitted because they vanish when projected to zero
momentum.}
\begin{tabular}{|c|cccc|}
\hline
$\o{B}$              & Disp. & $G_{1g}$ & $G_{2g}$ & $H_g$ \\
\hline
                     & $A_1$ &     8    &    0     &    4  \\
$\o{N}$              & $T_1$ &     8    &    3     &   11  \\
                     &  $E$  &     4    &    4     &   12  \\
\hline
                     & $A_1$ &     4    &    0     &    3  \\
$\o\Delta,\o\Omega^-$& $T_1$ &     4    &    1     &    5  \\
                     &  $E$  &     3    &    3     &    7  \\
\hline
                     & $A_1$ &     4    &    0     &    1  \\
$\o{\Lambda}$        & $T_1$ &     5    &    1     &    6  \\
                     &  $E$  &     1    &    1     &    5  \\
\hline
                     & $A_1$ &     4    &    0     &    3  \\
$\o{\Sigma},\o{\Xi}$ & $T_1$ &     7    &    3     &   10  \\
                     &  $E$  &     3    &    3     &    7  \\
\hline
\end{tabular}
\label{table:one-link}
\end{center}
\end{table}
The second line of Eq.~(\ref{eq:one-link-irreps}) defines new
coefficients for linear combinations of displaced three-quark
operators that realize the $A_1$, $T_1$, or $E$-type operators of
Eq.~(\ref{eq:irrep-displacements}). Table~\ref{table:one-link}
shows the numbers of positive-parity, one-link-displaced operators
for different baryons. For the one-link-displaced operators, there
are inequivalent operators for each of the two ways that
$I={1\over 2}$ can be formed from three light quarks, namely, the
mixed-antisymmetric (MA) and mixed-symmetric (MS) combinations.
Operators corresponding to total derivatives are excluded because
they vanish when projected to zero total momentum.

We construct matrices of correlation functions
\begin{align}
C&_{kk'}^{(\Lambda\lambda)}(t) \delta_{\Lambda\Lambda'}
\delta_{\lambda\lambda'}= c^{(\Lambda\lambda k)*}_{\a\b\c,\ell}
c^{(\Lambda'\lambda' k')}_{\mu\nu\rho,\ell '}
\times \nn\\
&\sum_{\bf x} \langle 0 \vert B_{\a\b\c}({\bf x},t)
\hat{d}^\dag_{\ell} \hat{d}_{\ell '} \o B_{\a'\!\b'\!\c'\!}({\bf
0},0) \vert 0 \rangle \gamma^4_{\a' \mu}\gamma^4_{\b'
\nu}\gamma^4_{\c' \rho}, \label{eq:matrix_correlation_function}
\end{align}
where displacement operator $\hat{d}_{\ell}$ applied to baryon
fields denotes a one-link-displaced baryon field for $\ell=\pm x,
\pm y, \pm z$ and a quasi-local baryon for $\ell=0$. Lattice
operators belonging to different irreps or different rows are
orthogonal because of the octahedral symmetry of the lattice.
Different embeddings of a given irrep and row provide sets of
operators that we used in calculating the matrix of correlation
functions. The $\gamma_4$ matrices are included in
Eq.~(\ref{eq:matrix_correlation_function}) in order to produce a
Hermitian matrix of correlation functions. For the operators used
in this work, the $\gamma_4$ matrices always reduce to a factor
${\cal P}^{(\Lambda)}_{k'}=\pm 1$ when they act on the three
``barred'' quark fields in $\o B^{(\Lambda \lambda)}_k$. This
factor is the $\rho$-parity defined in Ref.~\cite{Basak:2005ir}.
We write a matrix of correlation function for irrep $\Lambda$ and
row $\lambda$ as
\begin{equation}
C^{(\Lambda\lambda)}_{kk'}(t)= {\cal P}^{(\Lambda)}_{k'} \sum_{\bf x}
\langle 0 \vert B^{(\Lambda\lambda)}_{k}({\bf x},t)
\o B^{(\Lambda\lambda)}_{k'}({\bf 0},0) \vert 0 \rangle.
\label{eq:matrix_correlation_function2}
\end{equation}

\section{Charge conjugation}
\label{sec:charge-conjugation}

By inserting complete sets of particle states $\vert n \rangle$
and antiparticle states $\vert \o n \rangle$ in
Eq.~(\ref{eq:matrix_correlation_function2}),
and using translational invariance to
extract the dependence on ${\bf x}$ and $t$,
correlation functions are expanded as
\begin{align}
C^{(\Lambda\lambda)}_{kk'}(t) &= {\cal P}_{k'} \delta_{{\bf P},0}
\! \Big[ \sum_n\theta(t) \langle0 \vert B^{(\Lambda\lambda)}_k
\vert n\rangle
\langle n\vert \o B^{(\Lambda\lambda)}_{k'}\vert0\rangle e^{-E_n t} \nn \\
&- \sum_{\bar{n}}\theta(-t)\langle 0\vert \o
B^{(\Lambda\lambda)}_{k'}\vert\bar{n}\rangle \langle\bar{n}\vert
B^{(\Lambda\lambda)}_{k} \vert 0 \rangle e^{E_{\bar{n}}t} \Big].
\label{eq:Coft-1}
\end{align}
 Using the charge conjugation relations $\l| \bar{n} \r> =
{\cal C} \l| n \r> e^{i \phi}$, and the invariance of the vacuum
state under charge conjugation, the antiparticle contributions in
the $t<0$ part of Eq.~(\ref{eq:Coft-1}) are rewritten as
\begin{align}
& {\cal P}^{(\Lambda)}_{k'} \langle 0 \vert {\cal C} \o
B^{(\Lambda\lambda)}_{k'} {\cal C}^\dag {\cal C} \vert \bar n
\rangle \langle \bar n \vert {\cal C}^\dag {\cal C}
B^{(\Lambda\lambda)}_{k}
{\cal C}^\dag \vert 0 \rangle \nn\\
&= {\cal P}^{(\Lambda_c)}_{k'} c_{\a\b\c,\ell}^{(\Lambda_c
\lambda_c k)*} c_{\a'\b'\c',\ell '}^{(\Lambda_c\lambda_c k')}
\langle 0 \vert B_{\a\b\c} \hat{d}^\dag_{\ell} \vert n \rangle^*
\langle n \vert \hat{d}_{\ell '} \o B_{\!\a'\!\b'\!\c'} \vert 0
\rangle^*. \nn
\end{align} where ${\cal P}^{(\Lambda_c)}_{k'} = - {\cal
P}^{(\Lambda)}_{k'}$.
The relation between correlation functions with different parities
is then~\cite{Basak:2005aq},
\begin{equation}
C^{(\Lambda\lambda)}_{kk'}(t) = - C^{(\Lambda_c \lambda_c)*}_{kk'}(-t).
\label{eq:forward_backward_corr}
\end{equation}

Applying the temporal lattice boundary conditions
$C^{(\Lambda\lambda)}_{kk'}(-t) = \eta_t
C^{(\Lambda\lambda)}_{kk'}(T-t)$, with $\eta_t=+1$ for periodic
boundary conditions and $\eta_t = -1$ for antiperiodic boundary
conditions, the matrix of correlation functions can be written as
\begin{align}
&C^{(\Lambda \lambda)}_{kk'}(t) = \!
\delta_{{\bf P},0} \sum_n \! \Big[
\theta(t){\cal P \!}^{(\Lambda)}_{k'} \langle 0\vert B^{(\Lambda \lambda)\!}_k\vert n\rangle
\langle n\vert\o B^{(\Lambda \lambda)\!}_{k'}\vert0\rangle e^{\! -E_n t} \nn\\
-& \eta_t \theta(T \!\! - \! t){\cal P \!}^{(\!\Lambda_c \!)}_{k'}
\langle 0\vert B^{\! (\!\Lambda_c \!\lambda_c\!)}_k\vert n\rangle^{\!*} \!
\langle n\vert\o B^{(\Lambda_c \!\lambda_c\!)}_{k'}\vert0\rangle^{\!*}
e^{\!-E_{\bar{n}}\! (T-t)} \!\Big]\! . \!\!\!\!
\label{eq:Coft-3}
\end{align}
in the interval $0\le t < T$.
The forward propagating signal of a correlation function is equal to the
backward propagating signal of the parity-reversed, complex-conjugated
correlation function within the factor $-\eta_t$, i.e.,
\begin{equation}
C^{(\Lambda\lambda)}_{kk'}(t) = - \eta_t C^{(\Lambda_c \lambda_c)*}_{kk'}(T-t).
\label{eq:forward_backward_corr2}
\end{equation}

This symmetry is used to improve statistics. For each gauge
configuration we compute matrices of correlation functions using
for positive-parity operators the $\o B^{(\Lambda)}$-type
operators and for the corresponding negative-parity operators the
$\o B^{(\Lambda_c)}$-type operators. Starting with a matrix of
correlation functions for a given parity, a second matrix of
correlation functions is obtained for the opposite parity. After
time-reversal, complex conjugation and multiplication by the
factor $-\eta_t$, the second matrix of correlation functions is
averaged with the first one. A typical state created by the $\o
B^{(\Lambda\lambda)}_k$ operator has an effective mass plateau
occurring in first half of the time extent, $t< T/2$. Using the
$\o B^{(\Lambda_c\lambda_c)}_k$ operator, a similar plateau occurs
at $t>T/2$. These two plateaus are largely statistically
independent samples because gauge configurations on time slices
significantly before $T/2$ are largely uncorrelated with ones on time
slices significantly after $T/2$.

\section{Computational techniques}
\label{sec:computational_techniques}

\subsection{The variational method}

Analysis of excited state energies is based upon the matrices of
correlation functions defined in
Eq.~(\ref{eq:matrix_correlation_function2}).
Statistics are
improved by averaging over the rows of irreps, all of which are equivalent
because of the cubic symmetry. The matrix averaged over rows
is denoted as $C_{kk'}^{(\Lambda)}(t)$, i.e.,
the row label is omitted.

Because the operators used to form a matrix of correlation
functions are not normalized with respect to one another, we find
that they can produce diagonal elements of a correlation matrix
that differ by two orders of magnitude.  It is convenient to adopt
a normalization scheme at time $t_0$ such that each operator
produces a diagonal matrix element equal to 1. Thus, we define
normalization factors,
\begin{equation}
N_k(t_0) = \frac{1}{\sqrt{\l|C^{(\Lambda)}_{kk}(t_0)\r|}},
\label{eq:op_norm}
\end{equation}
and renormalize the operators by attaching a factor $N_k$ to
operator $B_k$.  The new matrix of correlation functions is
\begin{equation}
\widetilde{C}^{(\Lambda)}_{kk'}(t) = N_k
C^{(\Lambda)}_{kk'}(t) N_{k'},
\end{equation}
and its diagonal elements obey
$\widetilde{C}^{(\Lambda)}_{kk}(t_0) = 1$. Because of this
normalization convention the components of eigenvectors indicate
the relative importance of the contributions of various operators
to an eigenstate.

In order to extract an energy spectrum from the matrix of
correlation functions, we numerically solve the following
generalized eigenvalue equation,
\begin{equation}
\sum_{k'} \widetilde{C}^{(\Lambda)}_{kk'}(t) v^{(n)}_{k'}
(t,t_0) = \alpha^{(n)}(t,t_0) \sum_{k'}
\widetilde{C}^{(\Lambda)}_{kk'}(t_0) v^{(n)}_{k'}(t,t_0),
\label{eq:new_generalized_eigenvalue_equation}
\end{equation}
where superscript $n$ labels the eigenstates, such as the ground
state, the first excited state, and so forth. The reference time
$t_0$ in Eq.~(\ref{eq:new_generalized_eigenvalue_equation}) is
taken near the source time $t=0$ in order to have significant
contributions from excited states and to ensure stability of the
Cholesky procedure below. The principal eigenvalues
$\alpha^{(n)}(t,t_0)$ are related to the energy $E_n$
by~\cite{lw90}
\begin{equation}
\alpha^{(n)}(t,t_0) \simeq e^{-E_n(t-t_0)}\Big( 1 + {\cal O}
( e^{-|\delta
E|t}) \Big), \label{eq:generalized_eigenvalue}
\end{equation}
where $\delta E$ is the difference between $E_n$ and the next
closest energy.

 Energies $E_n$ are calculated from the
generalized eigenvalues according to
\begin{equation}
E_n = -\ln \l[ \alpha^{(n)} (t+1,t_0) \over \alpha^{(n)}(t,t_0) \r].
\end{equation}
We also perform fits of $\alpha^{(n)}(t,t_0)$ to
an exponential function
$\alpha^{(n)}(t,t_0) = e^{-E_n(t-t_0)}$ over a range
of time slices in order to better
determine the error of the energy $E_n$.

 Numerical solutions of the generalized
eigenvalue equation are obtained by performing a
Cholesky-decomposition of the matrix of correlation functions at
the reference time, $t_0$,
\begin{equation}
\widetilde{C}^{(\Lambda)}_{kk'}(t_0) = A^T_{kk''} A_{k''k'}.
\end{equation}
For the transfer matrix,
\begin{equation}
T (t,t_0) = (A^T)^{-1} \widetilde{C}^{(\Lambda)}(t) A^{-1}.
\end{equation}
the eigenvalue problem is
\begin{align}
T_{kk'}(t,t_0) V^{(n)}_{k'}(t,t_0) &= \alpha^{(n)}(t,t_0) V^{(n)}_k(t,t_0),
\label{eq:evalue_modified} \\
V^{(n)}_k (t,t_0) &= A_{kk'} v^{(n)}_{k'}(t,t_0).
\end{align}
Left and right eigenvectors of the transfer matrix
$T_{kk'}(t,t_0)$ are the same.

Eigenvectors in the generalized eigenvalue equation,
Eq.~(\ref{eq:new_generalized_eigenvalue_equation}), are orthogonal
with respect to $\widetilde{C}^{(\Lambda)}_{kk'}(t_0)$, {\it
i.e.},
\begin{equation}
v^{(n)T}_{k}(t,t_0) \widetilde{C}^{(\Lambda)}_{kk'}(t_0)
v^{(n')}_{k'}(t,t_0) = \delta_{nn'}. \label{eq:orthog_condition}
\end{equation}
It follows from
Eqs.~(\ref{eq:new_generalized_eigenvalue_equation}) and
(\ref{eq:orthog_condition}) that the correlation matrix at time
$t$ is diagonalized by these vectors, i.e.,
\begin{equation}
v^{(n)T}_{k}(t,t_0) \widetilde{C}^{(\Lambda)}_{kk'}(t)
v^{(n')}_{k'}(t,t_0) = \alpha^{(n)}(t,t_0) \delta_{nn'}.
\end{equation}
It is useful to define improved operators by first adopting a
suitable normalization of the eigenvector components. Let
\begin{equation}
\widetilde{v}^{(n)}_{k}(t,t_0) \equiv Z_n(t)N_k
v^{(n)}_{k}(t,t_0), \label{eq:Z_n}
\end{equation}
such that
\begin{equation}
\sum_k |\widetilde{v}^{(n)}_{k}(t,t_0)|^2 =  1. \label{eq:v-norm}
\end{equation}
The factor $Z_n$ allows a comparison of the overall normalization
of eigenvectors calculated on different volumes.

\subsection{Lattice setup and tuning}

A technical difficulty in extracting excited state energies from
lattice QCD simulations is that the signal-to-noise ratio degrades
for increasing time. A plateau in the energy may not be realized
for excitations. We use several techniques to overcome this
problem, namely, the optimization of operators so as to achieve
early plateaus of the energies, the imposition of symmetries and
the use of more gauge configurations to increase the statistics,
and the use of anisotropic lattices. Anisotropic lattices are
designed with temporal lattice spacing $a_t$ smaller than spatial
lattice spacing $a_s$. Using a finer lattice spacing along the
time direction provides more time-slices for analysis of
correlation functions at small time separations from the source.
Of course anisotropic lattices require tuning of coefficients in
the action.


In this work, anisotropic lattices with two different volumes are
used: 239 gauge field configurations on a $16^3\times 64$ lattice
and 167 configurations on a $24^3\times 64$ lattice. For both
lattices, the renormalized ratio of spatial lattice spacing to
temporal spacing is $\xi = a_s/a_t = 3.0$ and the temporal lattice
spacing corresponds to $a^{-1}_t = 6.0$ GeV~\cite{Basak:2004hr}.
Gauge-field configurations are generated using the anisotropic,
unimproved Wilson gauge action in the quenched approximation with
$\beta = 6.1$

\subsubsection{Anisotropic Wilson gauge action}
\label{subsec:anisotropic_wilson_gauge_action}

The anisotropic Wilson gauge
action~\cite{Chen:2000ej,Klassen:1998ua,Karsch:1997dc,Fujisaki:1996vv}
is given by
\begin{align}
S^\xi_G =& {\beta \over N_c} \Big[ {1\over\xi_0} \sum_{x,s>s'}
  {\rm Re Tr} \l( 1-P_{ss'}(x)\r) \nn\\
&+ \xi_0 \sum_{x,s} {\rm Re Tr} \l( 1-P_{st}(x) \r) \Big].
\end{align}
The renormalized anisotropy $\xi=3$ is held fixed in our
calculations and the bare anisotropy, $\xi_0$, is varied in order
to obtain the target value of $\xi$ for the desired value of
$\beta = 6.1$. Because the quenched approximation is used, tuning
of the gauge action may be performed independently of the fermion
action. Determination of the renormalized anisotropy is based on
measurements of static potentials for a quark-antiquark pair.

\subsubsection{Anisotropic Wilson fermion action}
\label{subsec:anisotropic_wilson_fermion_action}

The anisotropic Wilson fermion action
has the form,
\newcommand{\dslash}{D\hspace{-2.3mm}\slash\hspace{0.3mm}}
\begin{align}
S^\xi_F = a_t a_s^3 \sum_x \o{q} (x) \l[ m_0 + \nu_t \l( \gamma_4 \nabla_t-
{a_t \over 2} \Delta_t \r) \r. \nn\\
+ \nu_s \sum_s \l. \l(\gamma_s \nabla_s - {a_s\over 2}
\Delta_s \r) \r] q(x),
\label{eq:aniso_wilson_fermion_action}
\end{align}
where
\begin{align}
\Delta_\mu q(x) &= {U_\mu(x) q(x+\hat{\mu}) +
U^\dag_\mu(x-\hat{\mu}) q(x-\hat{\mu}) - 2 q(x) \over a_\mu^2}, \nn \\
\nabla_\mu q(x) &= {U_\mu(x) q(x+\hat{\mu}) -
U^\dag_\mu(x-\hat{\mu}) q(x-\hat{\mu}) \over 2a_\mu}.
\end{align}
Note that when $\nu_t=\nu_s$, the fermion action in
Eq.~(\ref{eq:aniso_wilson_fermion_action}) becomes the original
Wilson fermion action~\cite{Wilson:1974sk}. We hold $\nu_t = 1$ in
the tuning because that preserves the projection property of $(1
\pm \gamma_4)/2$ for parity. Then $\nu_s$ and the bare quark mass
$m_0$ are tuned in order to obtain the desired pion mass and the
correct speed of light.

\begin{figure}[t]
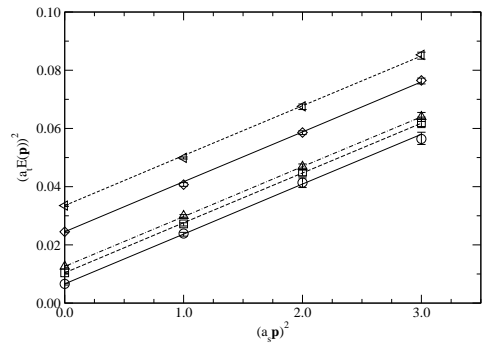

\hspace{0cm}
\ig[angle=0,width=0.35\textwidth]{mass16-prd-crop.eps}
\vspace{2mm}
 \caption{Pion dispersion relation based on tuned
parameters $m_0$ and $\nu_s$. The renormalized anisotropy is $3.0$
and the lattice volume is $16^3\times 64$ with $\beta=6.1$. Lines
show the the $c=1$ dispersion relation. The meaning of symbols is
given in Table~\ref{tab:nu_tuning}. } \label{fig:mass_nu}
\end{figure}

The tuning is based on the relativistic dispersion relation for
the pion, $E^2({\bf p}) = m^2c^4 + c^2 {\bf p}^2$, where $E({\bf
p})$ is the energy of a pion with total momentum ${\bf p}$ and
$mc^2=E(0)$. Expressing quantities in terms of the lattice
spacing, the speed of light is
\begin{equation}
c({\bf p}) = \xi \sqrt{ (a_t E({\bf p}))^2 - (a_t E(0))^2 \over
(a_s {\bf p})^2},
\end{equation}
where $\xi$ is the renormalized anisotropy. We measured the pion
energy $E({\bf p})$ for ${\bf p}=0$ and the lowest three nonzero
values of momentum, namely, ${\bf p} = {2\pi \over L} (1,0,0)$,
${2\pi \over L} (1,1,0)$, and ${2\pi \over L} (1,1,1)$.  We tuned
$\nu_s$ for five different bare quark masses $m_0$. The dispersion
relation for tuned parameter $\nu_s$ on the $16^3\times 64$
lattice is plotted in Fig.~\ref{fig:mass_nu}.
The straight lines passing
through the mean value of $E^2(0)$ show the desired continuum
dispersion relation for each pion mass.
In physical units, the volume is about $(1.6\, {\rm fm})^3 \times
(2.1\, {\rm fm})$.
\begin{table}[h]
\caption{Tuning parameters for the fermion action on the
$16^3\times 64$ lattice using
renormalized anisotropy $\xi=3$ and $\nu_t=1$}
 \label{table:tuning_params}
\begin{ruledtabular}
\begin{tabular}{lllll}
Symbol           & $a_t^2m_{\pi}^2 $ &~$m_{\pi}$ (MeV) &~$m_0$    &~$\nu_s$ \\
$\circ$          &  0.0066(5)        & 490             &  -0.313  &  0.898  \\
$\Box$           & 0.0104(5)         & 610             &  -0.318  &  0.901  \\
$\bigtriangleup$ &  0.0126(5)        & 675             &  -0.305  &  0.902  \\
$\Diamond$       &  0.0224(5)        & 900             &  -0.290  &  0.910  \\
$\lhd $          &  0.0335(5)        & 1100            &  -0.280  &  0.915
%
%
%
%
%
\label{tab:nu_tuning}
\end{tabular}
\end{ruledtabular}
\end{table}
For the anisotropic, unimproved Wilson fermion action, tuning of
the fermion action produced the results given in
Table~\ref{table:tuning_params}. A periodic boundary condition is
employed for spatial directions and an anti-periodic boundary
condition is employed for the temporal direction. Our spectrum
calculations are based on the parameter set that yields a pion
mass of $490$\,MeV.

Gauge links are smeared at source and sink according to the
APE-smearing method~\cite{Albanese87},
\begin{align}
&U^{(n+1)}_j(x) \rightarrow \nn\\
& U^{(n)}_j (x) + {1\over \alpha}
\sum_{k\perp j} U^{(n)}_k (x) U^{(n)}_j (x+\hat{k}) U_k^{\dag (n)} (x+\hat{j}),
\label{eq:ape_smearing}
\end{align}
where projection of gauge link matrices to SU(3) is performed
after each iteration. We used the APE-smearing parameters
$(\alpha, n)=(2.5, 3)$ throughout this work.

Quark fields are smeared for both quasi-local and
one-link-displaced operators using the Gaussian smearing
method~\cite{Alford:1995dm},
\begin{equation}
q_\mu ({x},t) \rightarrow \sum_{x'} \hat{G}^{(N)}({x},{x'}) q_\mu({x'},t),
\label{eq:GaussianSmearing}
\end{equation}
where $\hat{G}^{(N)}$ is an operator that acts recursively,
\begin{align}
\hat{G}^{(N)} ({x},{x'}) &= \sum_y ( \delta_{x,y}
+\sigma^2 \nabla_{x,y}^2 /4N )
\hat{G}^{(N-1)} ({y},{x'}),\nn \\
\hat{G}^{(0)} ({x},{x'}) &= \delta_{x,x'},
\end{align}
and $\nabla_{x,x'}^2$
is a three-dimensional, gauge-covariant Laplacian operator. We
used $(\sigma, N)=(3.0, 20)$ in this calculation.
%

\section{Discussion of results for each symmetry channel}
\label{sec:results}

We have extracted energies for $I={1\over 2}$ and $I={3\over
2}$ channels by diagonalizing matrices of correlation functions
formed from three-quark operators that share the same octahedral
symmetry. A large number of operators is available so that
reduction to a set of the most important operators is carried out
in the initial stage. The available operators are divided into
several subsets containing roughly 3 to 8 operators each and we
diagonalize the matrices computed from each subset of operators.
Inspection of the energies and corresponding eigenvectors reveals
which operators provide stronger signals for the low-lying states
of interest. In the second stage we omit all but these good
operators and form new matrices of correlation functions, which
are then diagonalized. This two-stage procedure yields solid
results for energies of low-lying states and is less susceptible
to noise than applying the variational method directly to the
matrices of largest dimension.

\begin{figure*}
\begin{tabular}{ccccc}
\ig[width=0.193\tw]{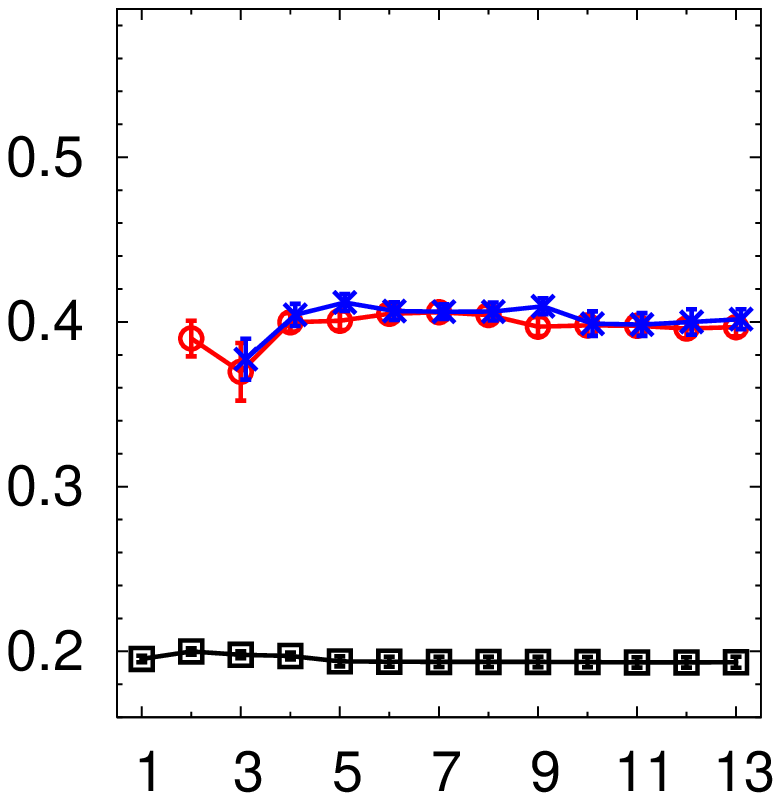} &
\ig[width=0.193\tw]{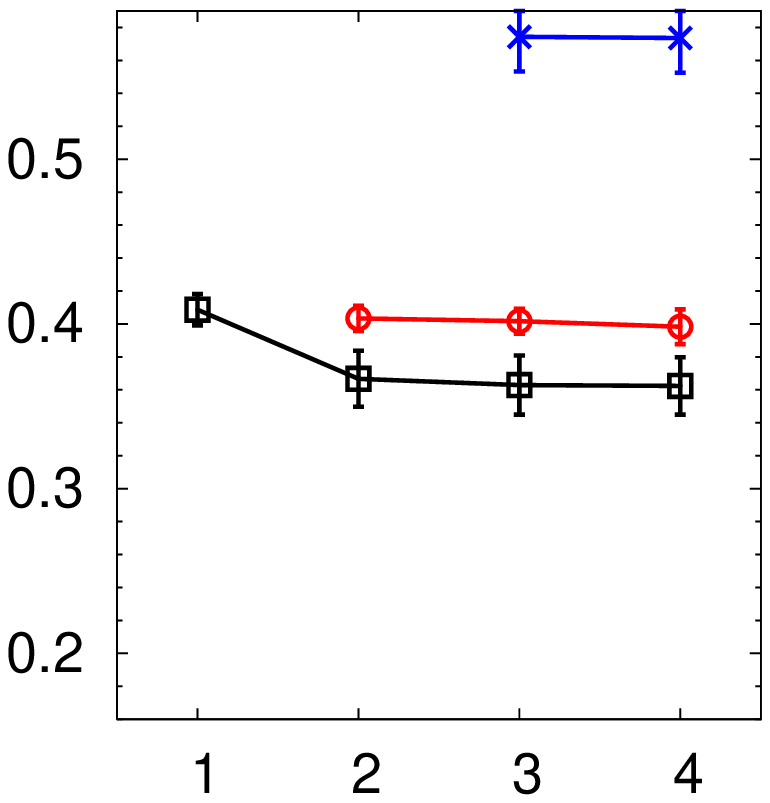} &
\ig[width=0.193\tw]{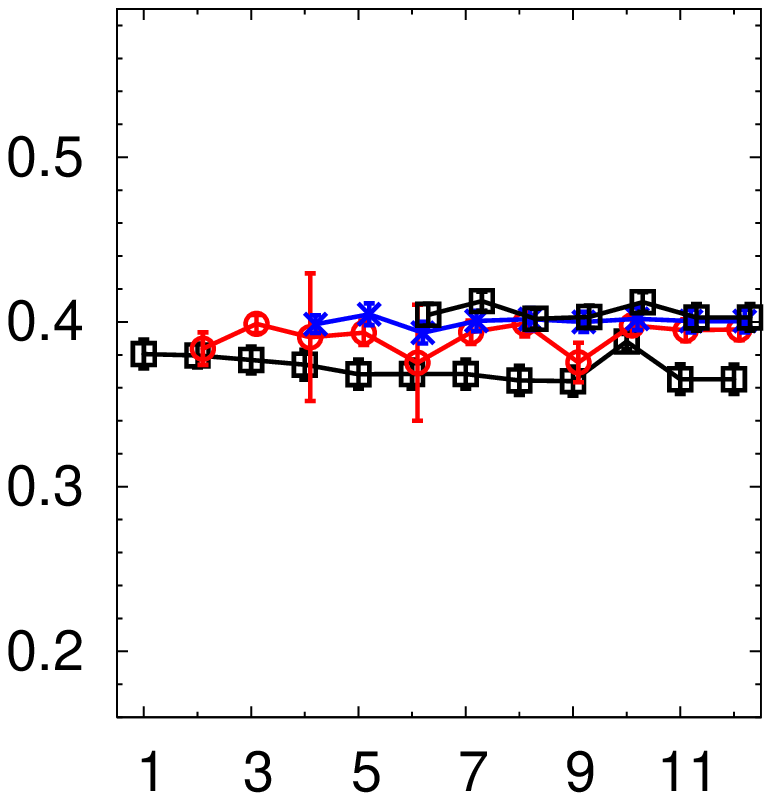}  &
\ig[width=0.193\tw]{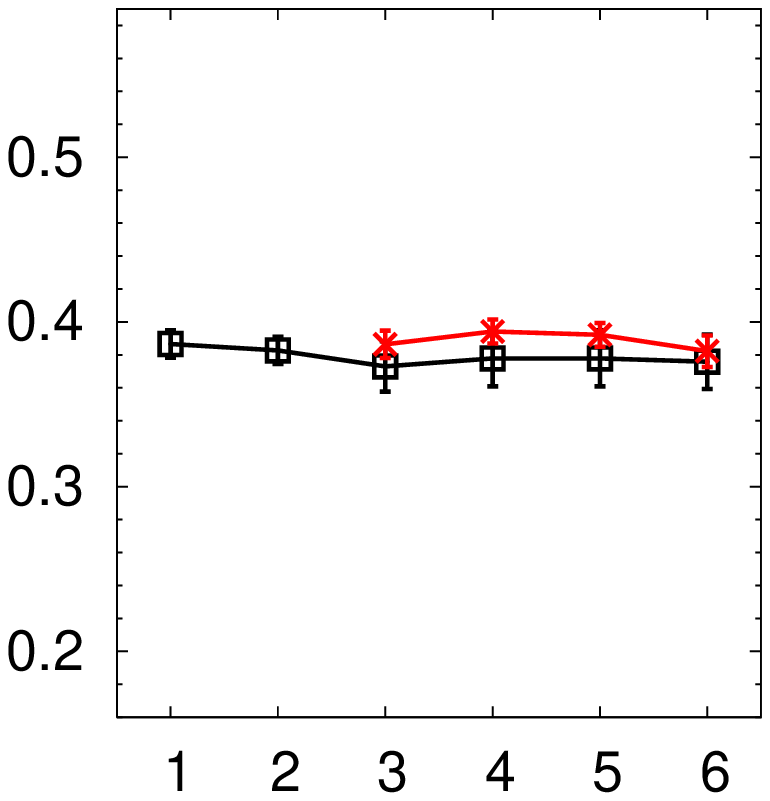} &
\ig[width=0.193\tw]{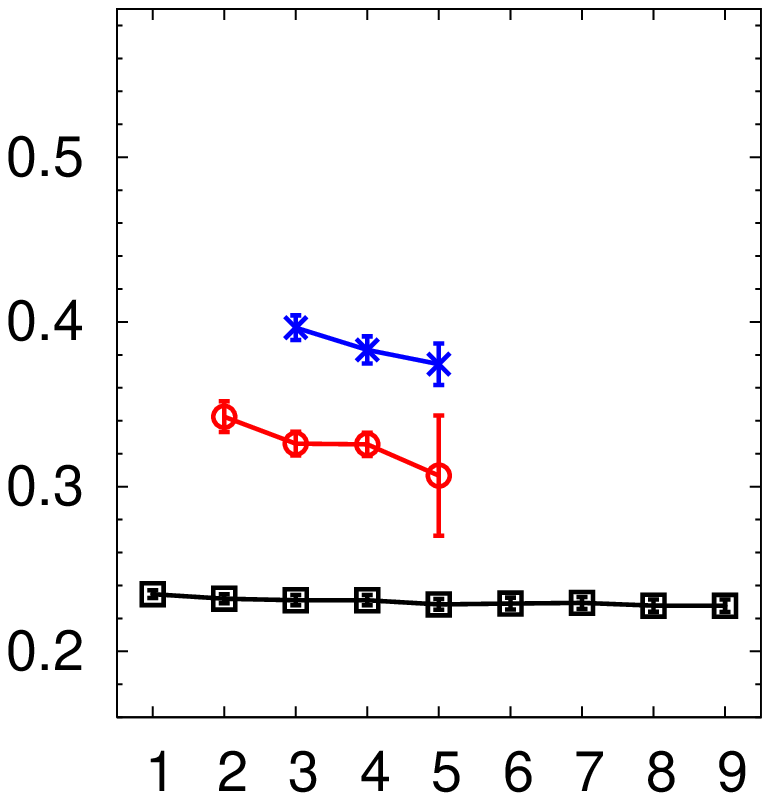}  \\
\ig[width=0.193\tw]{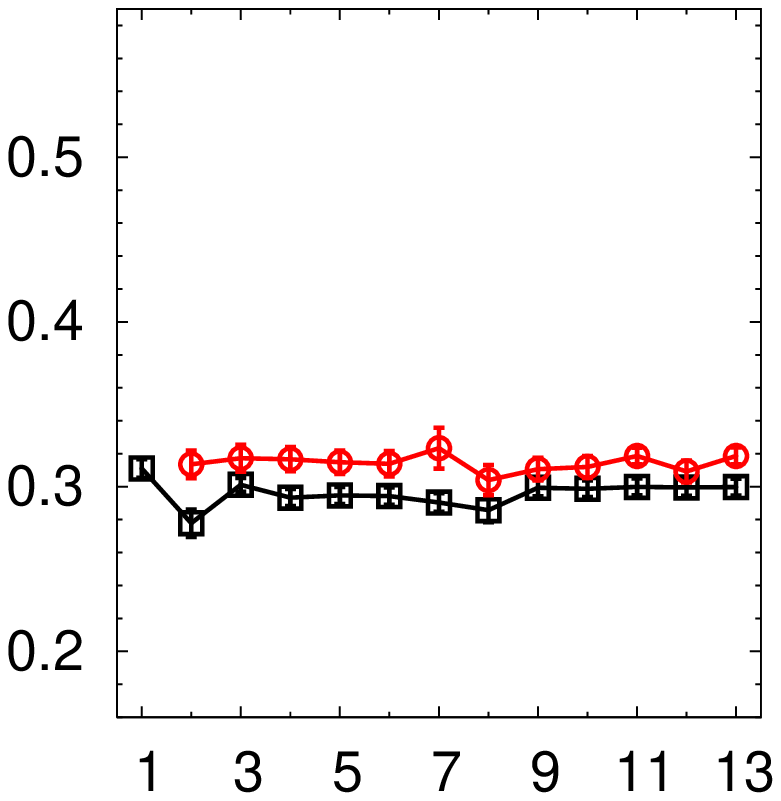} &
\ig[width=0.193\tw]{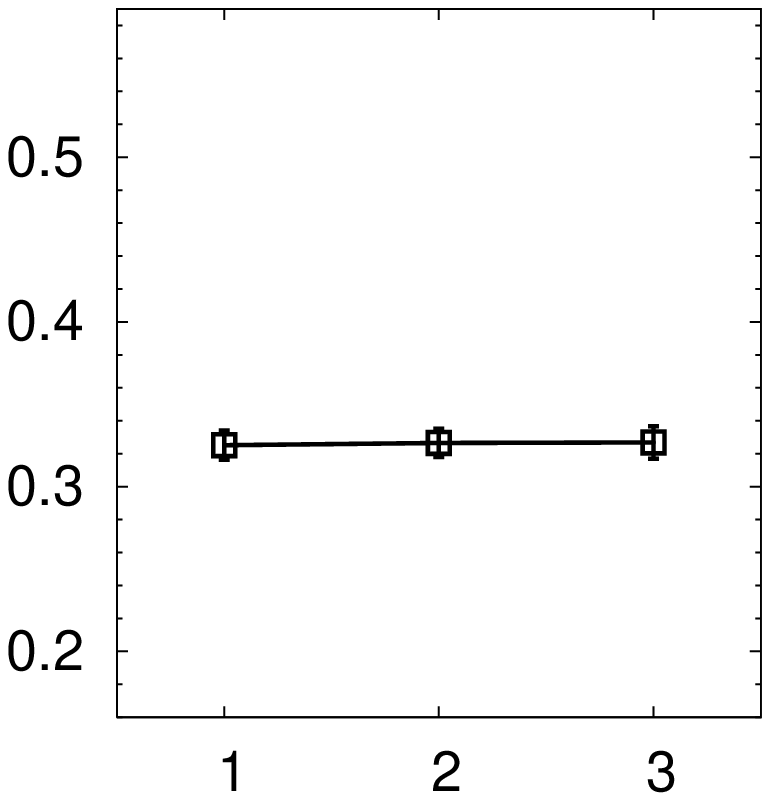} &
\ig[width=0.193\tw]{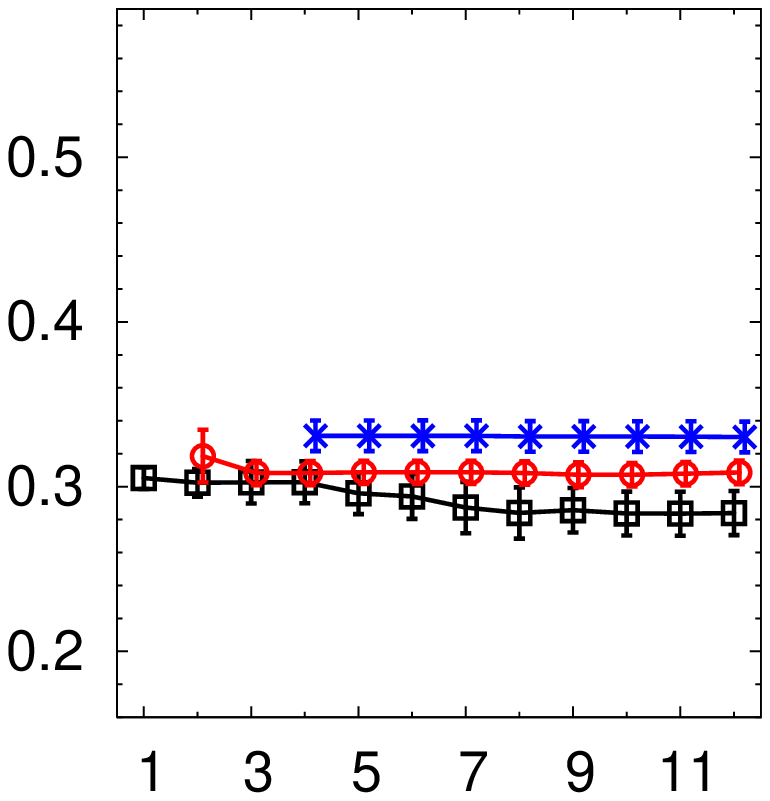}  &
\ig[width=0.193\tw]{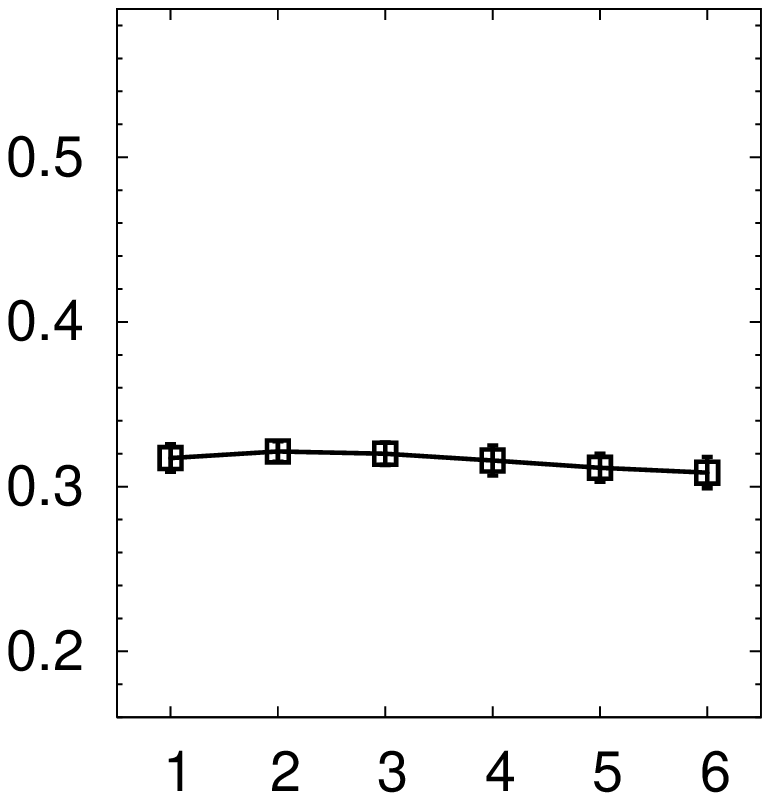} &
\ig[width=0.193\tw]{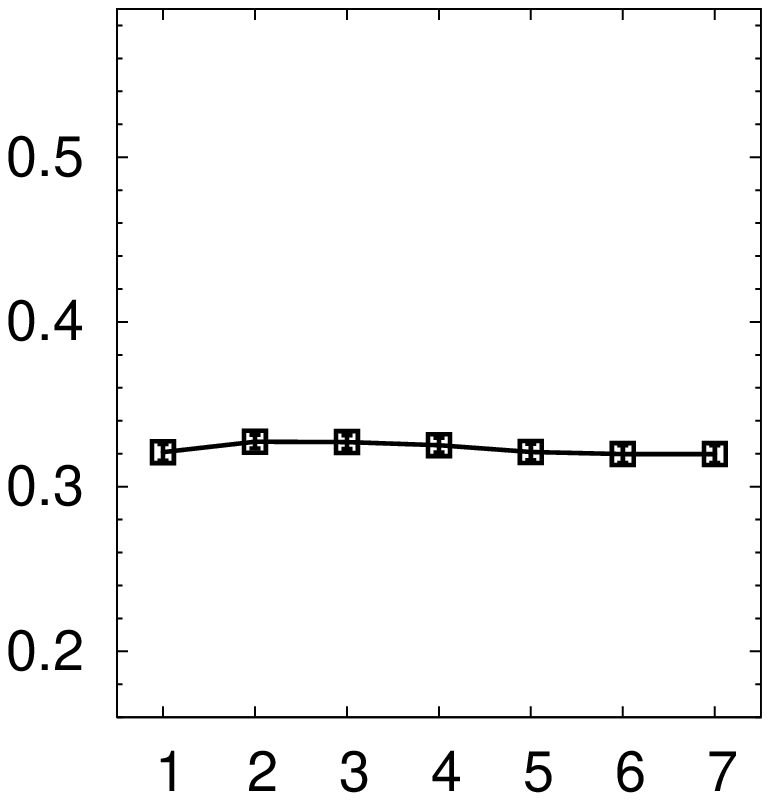}  \\
(a) $N, G_{1}$  &
(b) $N, G_{2}$  &
(c) $N, H$      &
(d) $\D, G_{1}$ &
(e) $\D, H$     \\
\end{tabular}
\caption{Energies in lattice units vs.\ dimension of the matrices
of correlation functions. Data for the $24^3\times64$ lattice are
used. Upper panels show energies for positive parity states and
lower panels show energies for negative parity states. The
$I=\frac{3}{2}, G_{2g/u}$ states are excluded because correlation
functions are based on a single operator.} \label{fig:stability}
\end{figure*}
In each symmetry channel, we have studied how the energies depend
upon the number of good operators that are used in applying the
variational method. Starting from a given set of good operators,
matrices of correlation functions of reduced dimension are
obtained by omitting a less important operator from the selected
set. Continuing this process, we have diagonalized matrices of
various dimensions and obtained energies. Low-lying energies in
each symmetry channel are plotted as a function of matrix
dimension in Fig.~\ref{fig:stability}. The $I=\frac{3}{2},
G_{2g/u}$ channel is not shown because results in that channel are
based on a single operator. The lowest energy states are more or
less stable as long as a few important operators are retained.
However, it is expected that use of more than the minimum number
of operators improves the lowest-lying states because it reduces
contamination from higher lying states. This expectation appears
to hold true for the lowest and first excited states but is less
evident for higher excited states with the limited set of
operators used in this work.

Diagonalization of a matrix of correlation functions provides the
principal eigenvalues $\alpha^{(n)}(t,t_0)$. These have been fit
with a single exponential form over a selected range of times,
$t_1 \le t \le t_2$ in order to extract energies, i.e.,
$\alpha^{(n)}(t,t_0) = C e^{-E_n(t-t_0)}$.  The resulting energies
are given in
Tables~\ref{table:Nstar_mass_gerade},~\ref{table:Nstar_mass_ungerade}
,~\ref{table:Delta_mass_gerade},
and~\ref{table:Delta_mass_ungerade}, which show the energies
obtained for positive-parity states with $I={1\over 2}$,
negative-parity states with $I={1\over 2}$, positive-parity
states with $I={3\over 2}$, and negative-parity states with
$I={3\over 2}$, respectively. Operators are selected so as
to provide the best quality of energies for the larger lattice
volume and the same sets of operators are then used for the
smaller lattice volume. The $Z$-factors given in
Eq.~(\ref{eq:Z_n}) are also shown along with the energies. Each
table shows results for two lattices, $16^3\times64$ and
$24^3\times64$, and six symmetry channels, $G_{1g/u}$, $G_{2g/u}$
and $H_{g/u}$. In some channels, there are states in the
$16^3\times64$ analysis that are missing from the $24^3\times64$
analysis and vice versa.

\renewcommand{\t}{\times}
\begin{table}[ht]
\caption{The $I = {1\over 2}$ $N^*$, positive-parity energies in
the units of $a_t^{-1}$ obtained on a $24^3\times64$ lattice
(left) and a $16^3\times64$ lattice (right). The irrep is shown in
the first row together with the dimension of the matrix of
correlation functions used. In the first column, energy values
obtained from fitting the generalized eigenvalues are shown in
increasing order, namely, the second row contains energies of the
lowest-lying states for each channel and the third row contains
the first excited state energies, {\it etc}. The Z-factor of
Eq.~(\ref{eq:Z_n}) is given in the second column. A time range in
which the energy and the Z-factor are fitted on the $24^3\times
64$ lattice is listed in the third column.}
\label{table:Nstar_mass_gerade}
\begin{ruledtabular}
\begin{tabular}{cc}
$G_{1g}$, $10\t10$, $24^3\t64$ &
$G_{1g}$, $10\t10$, $16^3\t64$ \\
\begin{tabular}{ccc}
\hline
energy   & $Z_n$       & time  \\
0.193(3) & 5.75(15)$\t10^{-5}$& 19-24 \\
0.398(6) & 4.79(12)$\t10^{-5}$&  8-12 \\
0.399(7) & 5.64(33)$\t10^{-5}$& 10-15
\end{tabular}
&
\begin{tabular}{cc}
\hline
energy   & $Z_n$        \\
0.194(4) & 8.45(15)$\t10^{-5}$ \\
0.403(6) & 9.53(26)$\t10^{-5}$  \\
0.409(8) & 8.84(46)$\t10^{-5}$
\end{tabular}
\\
$G_{2g}$, $4\t4$, $24^3\t64$ &
$G_{2g}$, $4\t4$, $16^3\t64$ \\
\begin{tabular}{ccc}
\hline
energy   & $Z_n$       & time  \\
0.377(13)& 1.47(57)$\t10^{-5}$&13-18 \\
0.398(10)& 1.34(25)$\t10^{-5}$&10-14 \\
0.574(2) & 3.73(30)$\t10^{-5}$&11-14
\end{tabular}
&
\begin{tabular}{cc}
\hline
energy   & $Z_n$           \\
0.414(13)& 2.11(35)$\t10^{-5}$ \\
0.444(7) & 7.66(100)$\t10^{-5}$  \\
0.564(12)& 5.90(199)$\t10^{-5}$
\end{tabular}
\\
$H_{g}$, $12\t12$, $24^3\t64$ &
$H_{g}$, $12\t12$, $16^3\t64$ \\
\begin{tabular}{ccc}
\hline
energy   & $Z_n$       & time  \\
0.365(9) & 5.40(15)$\t10^{-5}$&13-18 \\
0.395(7) & 5.14(14)$\t10^{-5}$&10-15 \\
0.400(7) & 4.39(7)$\t10^{-5}$&10-16 \\
0.403(8) & 4.27(10)$\t10^{-5}$&10-14 \\
         &             &       \\
         &             &
\end{tabular}
&
\begin{tabular}{cc}
\hline
energy   & $Z_n$         \\
0.375(9) & 9.25(33)$\times 10^{-5}$  \\
0.407(5) & 8.38(21)$\times 10^{-5}$  \\
0.407(8) & 7.75(10)$\times 10^{-5}$  \\
0.422(6) & 8.07(17)$\times 10^{-5}$  \\
0.427(8) & 6.36(8)$\times 10^{-5}$    \\
0.443(6) & 8.75(7)$\times 10^{-5}$
\end{tabular}
\end{tabular}
\end{ruledtabular}
\end{table}
\normalsize
\begin{table}[ht]
\caption{The $I = {1\over 2}$ $N^*$, negative-parity energies
in the units of $a_t^{-1}$
obtained on a $24^3\times64$ lattice
(left) and a $16^3\times64$ lattice (right).
A similar description applies as in the caption of
Table~\ref{table:Nstar_mass_gerade}.}
\label{table:Nstar_mass_ungerade}
\begin{ruledtabular}
\begin{tabular}{cc}
$G_{1u}$, $7\t7$, $24^3\t64$ &
$G_{1u}$, $7\t7$, $16^3\t64$ \\
\begin{tabular}{ccc}
\hline
energy   & $Z_n$       & time  \\
0.290(5) & 5.68(53)$\times 10^{-5}$ & 12-17 \\
0.323(12)& 5.73(44)$\times 10^{-5}$ & 11-17
\end{tabular}
&
\begin{tabular}{cc}
\hline
energy   & $Z_n$       \\
0.290(7) & 9.19(40)$\times 10^{-5}$ \\
0.340(6) & 9.35(73)$\times 10^{-5}$
\end{tabular}
\\
$G_{2u}$, $3\t3$, $24^3\t64$ &
$G_{2u}$, $3\t3$, $16^3\t64$ \\
\begin{tabular}{ccc}
\hline
energy   & $Z_n$       & time  \\
0.327(10)& 1.93(82)$\times 10^{-5}$ & 15-20 \\
0.473(11)& 7.24(22)$\times 10^{-6}$ &  7-12
\end{tabular}
&
\begin{tabular}{cc}
\hline
energy   & $Z_n$         \\
0.325(9) & 2.67(121)$\times 10^{-5}$  \\
         &
\end{tabular}
\\
$H_{u}$, $8\t8$, $24^3\t64$ &
$H_{u}$, $8\t8$, $16^3\t64$ \\
\begin{tabular}{ccc}
\hline
energy   & $Z_n$       & time  \\
0.284(16)& 5.00(38)$\times 10^{-5}$ & 18-23 \\
0.308(7) & 5.74(8)$\times 10^{-5}$  & 15-22 \\
0.330(9) & 7.93(5)$\times 10^{-5}$  & 16-22 \\
\end{tabular}
&
\begin{tabular}{cc}
\hline
energy   & $Z_n$       \\
0.284(9) & 8.07(26)$\times 10^{-5}$  \\
0.310(8) & 1.04(1)$\times 10^{-4}$  \\
0.328(8) & 1.44(1)$\times 10^{-5}$
\end{tabular}
\end{tabular}
\end{ruledtabular}
\end{table}
\begin{table}[ht]
\caption{The $I= {3\over 2}$ $\Delta^*$, positive-parity energies
in the units of $a_t^{-1}$
obtained on the $24^3\times 64$
lattice (left) and the $16^3\times64$ lattice (right).
A similar description applies as in the caption of
Table~\ref{table:Nstar_mass_gerade}.}
\label{table:Delta_mass_gerade}
\begin{ruledtabular}
\begin{tabular}{cc}
$G_{1g}$, $5\t5$, $24^3\t64$ &
$G_{1g}$, $5\t5$, $16^3\t64$ \\
\begin{tabular}{ccc}
\hline
energy   & $Z_n$        & time  \\
0.378(17)& 5.05(14)$\times 10^{-5}$  & 14-18 \\
0.392(7) & 7.41(26)$\times 10^{-5}$  & 12-17
\end{tabular}
&
\begin{tabular}{cc}
\hline
energy   & $Z_n$         \\
0.373(8) & 1.67(5)$\times 10^{-4}$   \\
0.401(13)& 9.33(16)$\times 10^{-5}$
\end{tabular}
\\
$G_{2g}$, $1\t1$, $24^3\t64$ &
$G_{2g}$, $1\t1$, $16^3\t64$ \\
\begin{tabular}{ccc}
\hline
energy   & $Z_n$       & time \\
0.416(7) & 8.92(28)$\times 10^{-5}$ & 9-14
\end{tabular}
&
\begin{tabular}{cc}
\hline
energy   & $Z_n$        \\
0.429(6) & 1.49(4)$\times 10^{-4}$
\end{tabular}
\\
$H_{g}$, $4\t4$, $24^3\t64$ &
$H_{g}$, $4\t4$, $16^3\t64$ \\
\begin{tabular}{ccc}
\hline
energy   & $Z_n$       & time  \\
0.231(3) & 1.22(3)$\times 10^{-4}$  & 19-26 \\
0.326(7) & 5.89(5)$\times 10^{-5}$  & 13-20 \\
0.370(25)& 6.54(4)$\times 10^{-5}$  & 19-23 \\
0.400(8) & 7.15(5)$\times 10^{-5}$  & 14-19
\end{tabular}
&
\begin{tabular}{cc}
\hline
energy   & $Z_n$         \\
0.232(4) & 1.79(4)$\times 10^{-4}$  \\
0.341(13)& 1.11(1)$\times 10^{-4}$   \\
0.399(10)& 1.20(6)$\times 10^{-4}$   \\
0.402(6) & 1.29(2)$\times 10^{-4}$
\end{tabular}
\end{tabular}
\end{ruledtabular}
\end{table}
\begin{table}[ht]
\caption{The $I= {3\over 2}$ $\Delta^*$, negative-parity energies
in the units of $a_t^{-1}$
obtained on the $24^3\times 64$
lattice (left) and the $16^3\times64$ lattice (right).
A similar description applies as in the caption of
Table~\ref{table:Nstar_mass_gerade}.}
\label{table:Delta_mass_ungerade}
\begin{ruledtabular}
\begin{tabular}{cc}
$G_{1u}$, $6\t6$, $24^3\t64$ &
$G_{1u}$, $6\t6$, $16^3\t64$ \\
\begin{tabular}{ccc}
\hline
energy   & $Z_n$       & time \\
0.309(9) & 3.13(99)$\times 10^{-6}$ & 15-20 \\
\end{tabular}
&
\begin{tabular}{cc}
\hline
energy   & $Z_n$          \\
0.312(7) & 5.38(117)$\times 10^{-5}$
\end{tabular}
\\
$G_{2u}$, $1\t1$, $24^3\t64$ &
$G_{2u}$, $1\t1$, $16^3\t64$ \\
\begin{tabular}{ccc}
\hline
energy   & $Z_n$       & time \\
0.506(7) & 7.31(16)$\times 10^{-5}$ & 6-12
\end{tabular}
&
\begin{tabular}{cc}
\hline
energy   & $Z_n$      \\
0.539(5) & 1.43(2)$\times 10^{-4}$
\end{tabular}
\\
$H_{u}$, $7\t7$, $24^3\t64$ &
$H_{u}$, $7\t7$, $16^3\t64$ \\
\begin{tabular}{ccc}
\hline
energy   & $Z_n$        & time  \\
0.320(5) & 8.38(30)$\times 10^{-5}$  & 12-17 \\
0.485(7) & 5.56(53)$\times 10^{-5}$  & 7-12
\end{tabular}
&
\begin{tabular}{cc}
\hline
energy   & $Z_n$      \\
0.317(6) & 1.16(4)$\times 10^{-4}$   \\
0.485(8) & 9.14(28)$\times 10^{-5}$
\end{tabular}
\end{tabular}
\end{ruledtabular}
\end{table}

We have examined the Z-factors defined in Eqs.~(\ref{eq:Z_n}) and
(\ref{eq:v-norm}) as a possible means to detect scattering states
following the methods of Ref.~\cite{Mathur:2004}.  This has not
produced clearly interpretable results. Therefore we discuss the
results for each channel in the following subsections assuming
that there are no scattering states in the extracted spectra for
the pion mass used in these calculations.

\subsection{The $G_1$ channel}

\begin{figure*}
\begin{tabular}{cc}
\ig[width=0.41\tw]{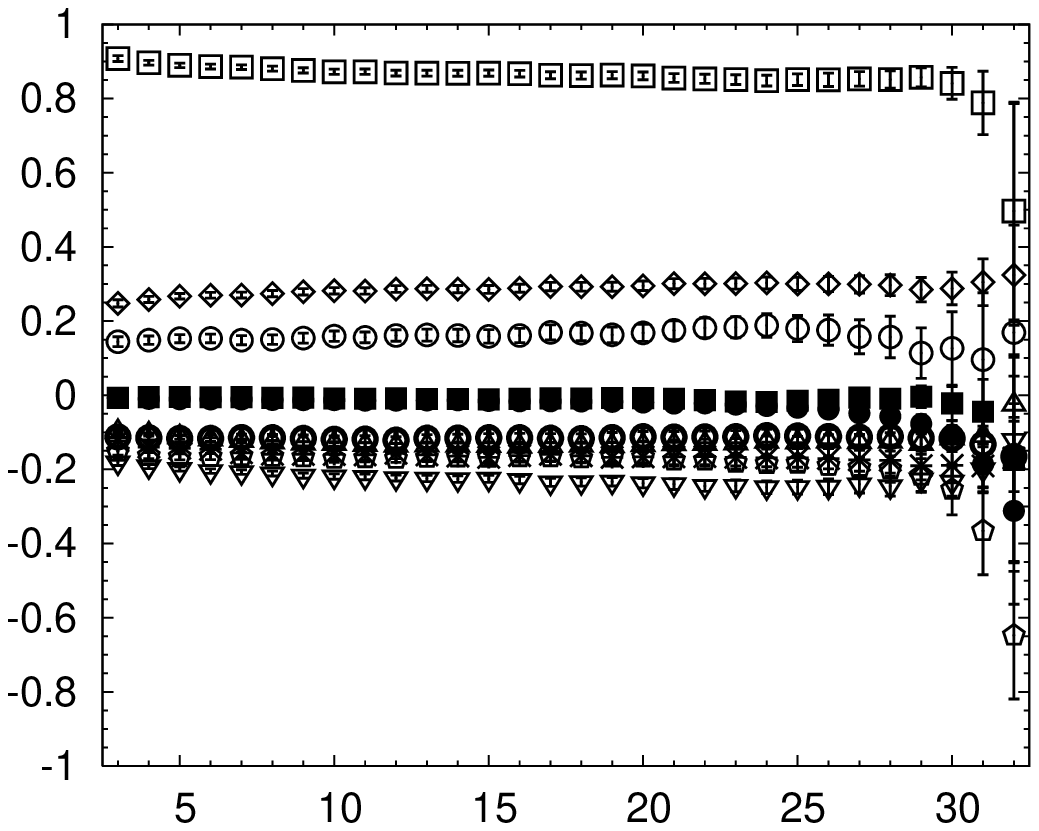} &
\ig[width=0.41\tw]{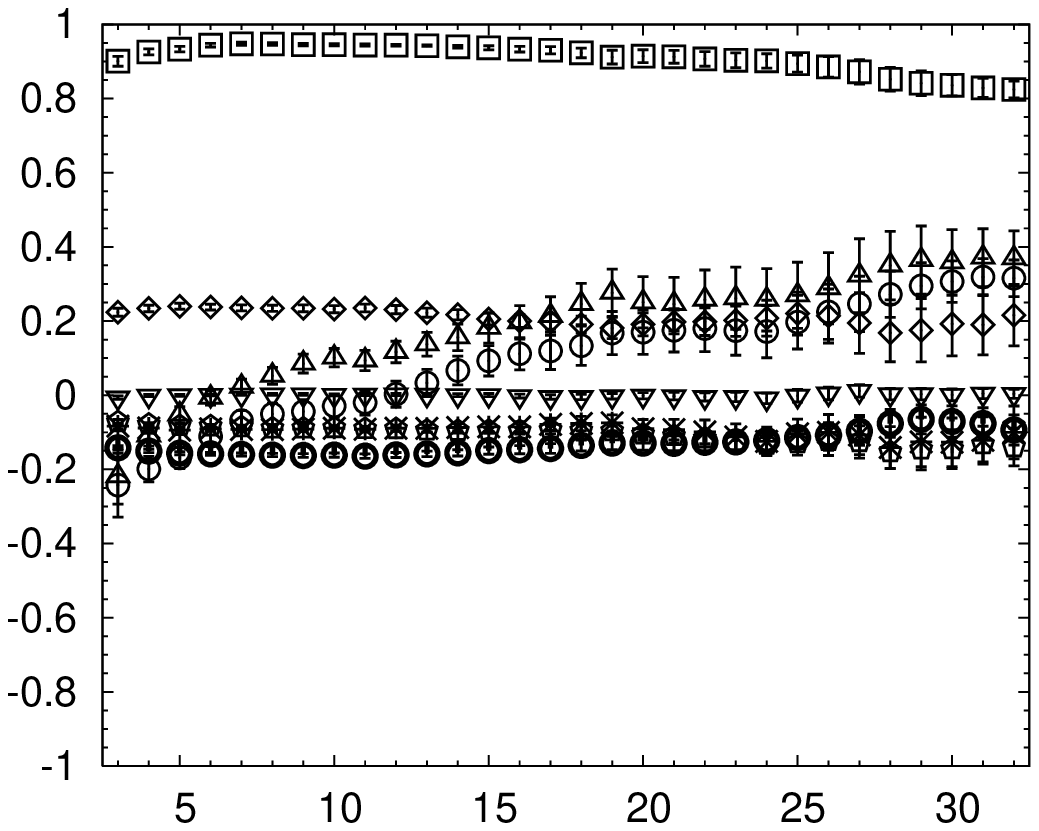} \\
\begin{tabular}{ccc}
\hline
$\Box$     & $\o N^{G_{1g},1}_{\frac{1}{2},\frac{1}{2}}$ & MA \\
$\Diamond$ &
${1\over\sqrt{2}}\hat{D}^- \o N^{H_u}_{\frac{3}{2},\frac{3}{2}}
-{1\over\sqrt{3}}\hat{D}^0 \o N^{H_u}_{\frac{3}{2},\frac{1}{2}}
+{1\over\sqrt{6}}\hat{D}^+ \o N^{H_u}_{\frac{3}{2},-\frac{1}{2}}$ & S \\
$\bigtriangledown$ &
$\sqrt{2\over 3}\hat{D}^+ \o N^{G_{1u},1}_{\frac{1}{2},-\frac{1}{2}}
-{1\over\sqrt{3}}\hat{D}^0 \o N^{G_{1u},1}_{\frac{1}{2},\frac{1}{2}}$ & MA \\
$\bigcirc$ &
$\o N^{G_{1g},2}_{\frac{1}{2},\frac{1}{2}}$ & MA \\
\hline
\end{tabular}
&
\begin{tabular}{ccc}
\hline
$\Box$ & $\o N^{H_u}_{\frac{3}{2}, \frac{3}{2}}$ & MA \\
$\bigtriangleup$ &
$\sqrt{3\over5} \hat{D}^0 \o N^{H_g}_{\frac{3}{2},\frac{3}{2}}
-\sqrt{2\over5} \hat{D}^+ \o N^{H_g}_{\frac{3}{2},\frac{1}{2}}$ & S \\
$\Diamond$ &
$\sqrt{3\over5} \hat{D}^0 \o N^{H_g}_{\frac{3}{2},\frac{3}{2}}
-\sqrt{2\over5} \hat{D}^+ \o N^{H_g}_{\frac{3}{2},\frac{1}{2}}$ & MA \\
$\bigcirc$ &
$\hat{D}^+ \o N^{G_{1g},1}_{\frac{1}{2},\frac{1}{2}}$ & MA \\
\hline
\end{tabular}
\\
(a) Nucleon, ground state &
(b) $N^*$, lowest $H_u$ state
\end{tabular}
\caption{Eigenvector components $v^{(0)}_k(t,0)$ of the ground state for
(a) the $I=\frac{1}{2}, G_{1g}$ channel using a $10\times 10$ matrix and
(b) the $I=\frac{1}{2}, H_u$ channel using an $8\times 8$ matrix, both
on the $24^3\times64$ lattice.
The vertical axis is the value of the coefficient of a basis operator and the
horizontal axis is time.
Each basis operator corresponds to a symbol in the plot; the four
most important operators are listed below the plot together with their symbols.
$\o N^{\lambda k}_{S, S_z}$ is our notation for a smeared three-quark operator
with spin $S$, spin projection $S_z$, irrep
$\Lambda$, and embedding $k$.  The symmetry (MA, MS or S) of Dirac indices
for each three-quark operator is written next
to the operator.
Explicit forms of the operators $\o N^{\lambda k}_{S, S_z}$ are
provided in Ref.~\cite{Basak:2005ir}.}
\label{fig:NG1g_eigenvectors}
\end{figure*}
In the $I=\frac{1}{2}, G_{1g}$ channel, we selected 10
operators on the $24^3\times 64$ lattice that give the best
quality of energies using the variational method, and the same 10
operators are used to analyze the $16^3\times 64$ lattice data.
The ground state of this channel, i.e., the nucleon state, has a
very stable set of eigenvector components with respect to time as
shown in Fig.~\ref{fig:NG1g_eigenvectors}. As many studies have
shown, the ground state is dominated by a local operator of the
form, $u^T(C\gamma_5)d P_+u$, where $P_\pm$ are parity projection
matrices for Dirac spinors. However, we find non-negligible
contributions from nonlocal $T_1$ operators represented by
diamonds, circles and inverted triangles in the figure. One
significant operator has a mixed-symmetric isospin symmetry with
totally symmetric Dirac indices. Although this isospin symmetry is
unusual, nonlocal operators formed from three quarks allow such
combinations. In fact, this operator plays a significant role in
the first excited state in combination with a local operator of
the second embedding. The first and second excited states have
relatively clean signals and are dominated by local and
$T_1$-displaced operators, respectively. Similar energies are
calculated for states on the two lattice volumes except for the
second excited state. Dominance of a nonlocal operator in the
second-excited-state eigenvector might explain this dependence on
lattice volume because we find that the nonlocal operators
generally exhibit more sensitivity to the spatial volume.
Eigenvectors are more or less the same in each state on the two
lattices as far as the coefficients of the two or three most
important operators are concerned. Fluctuations of the eigenvector
components generally increase with time for excited states, but
they are fairly stable in the time ranges where energies are
extracted.

In the $G_{1u}$ channel we obtain two states which could
correspond to $N({1\over 2}^-,1535)$ and $N({1\over 2}^-,1650)$.
Each is dominated by two local operators but in different linear
combinations. Contributions from nonlocal operators for the first
excited state are around $|v^{(n)}_k(t,0)|\simle 0.3$, which is
less than for positive parity states. The energy of the lowest
state on the two lattce volumes is essentially the same,
suggesting that a $16^3$ lattice is large enough to contain this
state with a 490\,MeV pion mass.

In the $I=\frac{3}{2}, G_{1g}$ channel, the first excited
(the lowest) state in $16^3$ volume corresponds to the lowest (the
first excited) state in $24^3$ volume.  This correspondence is
based on the similarity of eigenvector components. The first
excited state of $I=\frac{3}{2}, G_{1g}$ has lattice
energy of $0.392(7)$ on the larger volume and $0.373(8)$ on the
smaller volume, with error bars that do not overlap. The apparent
crossover of energies of the same state at different lattice
volumes is observed only in this particular state.
The eigenvectors for these states show that they are
dominated by $T_1$ displaced operators with MS isospin.

In the negative parity channel, it is more difficult to determine
the excited state energies due to the large error bars, while the
lowest state, corresponding to $\D({1\over 2}^-,1620)$, is fairly
stable. Again the $T_1$ displaced operators dominate the
eigenvectors of the two lowest energy states.

\subsection{The $H$ channel}

We find a richer spectrum in $I=\frac{1}{2}, H_{g/u}$ than
in other channels as shown in
Tables~\ref{table:Nstar_mass_gerade},~\ref{table:Nstar_mass_ungerade},
and~\ref{table:Delta_mass_gerade}. Plateaus are achieved for $t
\approx 20$ in the first three states. Approximate plateaus could
be selected for the next three states. However, the use of more
diverse operators together with better statistics is required in
order to extract reliably more than the first three states.

We obtained four energies for the $I={1\over 2}, H_g$ channel
using the $24^3$ lattice and six energies using the $16^3$
lattice, with errors relatively small compared with those in other
channels. A quasi-local operator yields the dominant coupling to
the lowest $H_g$ state and our operator is essentially the same as
the Rarita-Schwinger projected operator that has been used in
other works. The Rarita-Schwinger projected lattice interpolating
field belongs to the $H$ irrep and is therefore consistent with
the subduction of continuum spins ${3\over 2}, {5\over 2}, {7\over
2}, \cdots$. Several
works~\cite{Zhou:2006xe,Sasaki:2005uq,Sasaki:2005ug,Zanotti:2003fx}
have suggested that this lowest $H_g$ state likely corresponds to
a continuum spin ${3\over 2}^+$ state. However, we find an
indication that it corresponds to a higher spin, either ${5\over
2}^+$ as in the $N({5\over 2}^+,1680)$, or $\frac{7}{2}^+$. The
possible spins will be discussed in the next section when the
patterns of states over all the irreps are considered.

For states above the ground state, combinations of nonlocal $T_1$
operators dominate the couplings. Consistent with this, we observe
a finite volume effect on some energies in the $N^*$, $H_g$
excited spectrum. Even though the lowest energy state is dominated
by the quasi-local operator, its energy has some dependence on the
volume of the lattice. The lowest and the first excited states are
clearly identifiable from eigenvector compositions in both
volumes. Eigenvectors for the second and the third excited states
are contaminated by noise on the $24^3$ lattice with the available
statistics, although the $16^3$ lattice provides stable
eigenvectors.

The $I=\frac{1}{2}, H_u$ states provide the least
contaminated group of states in the sense that energies for
several excited states are determined easily and the lowest few
states have eigenvectors that are nearly as stable as those of the
nucleon $G_{1g}$ ground state. The lowest-lying state should
correspond to the physical $N({3\over 2}^-,1520)$ state, the first
excited state should correspond to the $N({5\over 2}^-,1675)$
state, and the second excited state should correspond to the
$N({3\over 2}^-,1700)$ state, supposing no scattering states are
involved. Energy differences between these states are relatively
small and our lattice results agree with this pattern. In contrast
to the positive parity channel, finite volume effects in energies
are well within statistical fluctuations.  The nonlocal
three-quark operators that we use yield nonvanishing MS isospin
combinations although the simplest quark model does not have them.
Our results show that contributions of MS isospin operators are
crucial to the first and second $H_u$ excited states. The same
operators are important in both volumes. Energy differences
between parity-partner states are found to be smaller for the
$I=\frac{1}{2}, H$ irreps than for other channels.

The $\Delta$ baryon appears as the lowest energy state in the
$I=\frac{3}{2}, H_g$ channel. A local operator with
nonrelativistic spin components dominates the ground state with
very small errorbars. The first and second excited state signals
should correspond to $\Delta({3\over 2}^+,1600)$ and
$\Delta({5\over 2}^+,1905)$, respectively. Signals for these
states are also reasonably clean on the $16^3$ lattice, however
the results on the $24^3$ lattice are subject to larger
fluctuations and the extracted energies are not reliable.

In the $I=\frac{3}{2}, H_u$ spectrum we find that the ground
state, corresponding to $\Delta({3\over 2}^-,1700)$, has stable
signatures in both energy and eigenvector plots but the signal is
short-lived. This is associated with the fact that the most
important operator is the negative parity transform of the
operator that dominates the positive parity ground state.  The
resulting long-lived plateau in the backward-in-time part of the
correlation function limits the range of times where the excited
state energies can be extracted. The energy differences between
parity-partner states of the low-lying delta baryon spectrum in
the $H$ irrep are larger than in other channels.

\subsection{The $G_2$ channel}

The $G_2$ baryon spectrum is the least explored because nonlocal
operators have not been readily available.  However, $G_2$
operators are very important for the assignment of higher spins
because of their role in the patterns of subduction of continuum
spins to the octahedral irreps.

We have selected the four best operators for the
$I=\frac{1}{2}, G_{2g}$ spectrum by the method described earlier.
One of these is the operator with $E$-type spatial displacement,
which transforms evenly under spatial inversion. We find that
$G_2$ energies are the most difficult to extract in our
calculations because the plateaus are relatively short-lived,
signal-to-noise ratios are small, and the number of operators that
we have used for the variational method is limited.  The
limitation on the number of operators arises because we have
restricted the operators to ones that can be constructed from
one-link displacements. Work in progress includes more varied
types of operators, which yield larger numbers of $G_2$
operators.~\cite{Lichtl-Thesis}

The lowest state in the $I=\frac{1}{2}, G_{2g}$ channel has
significant couplings to the $T_1$-displaced MA isospin operator,
the $T_1$-displaced MS isospin operator, and the $E$-displaced MS
isospin operator. There is a non-negligible contribution to the
lowest energy state of $G_2$ from the $E$-type of displaced
operator that corresponds to $L \ge 2$ in the subduction of
continuum angular momenta. The first excited state has significant
couplings to a combination of two $T_1$ operators that also
contribute to the ground state but in a different linear
combination. The second excited state couples almost purely to a
single $T_1$ operator with MS isospin. Eigenvectors are similar on
the two lattice volumes for the lowest three states, however their
energies change significantly. The lowest energy is calculated to
be $0.362(17)$ in this channel, which agrees very well with the
lowest $I=\frac{1}{2}, H_g$ energy, $0.365(9)$. Because
these two energies are degenerate within errors, they are
consistent with being the partner states for spin ${5\over 2}$.
However, that is not the only possibility as is discussed in the
next section.

For the $I=\frac{1}{2}, G_{2u}$ channel, the $T_1$-displaced
operator with MS isospin dominates the lowest state, while the
$E$-displaced operator plays a significant role with
$|v^{(n)}_k(t,0)|\sim 0.4$ in the first excited state. The lowest
physical state corresponding to $I=\frac{1}{2}, G_{2u}$ is
$N({5\over 2}^-,1675)$, which should also occur as the first
excited state of the $H_u$ spectrum. The lowest energy state of
$H_u$ should correspond to $N({3\over 2}^-,1520)$. We find that
the first excited state of the $H_u$ channel has energy
$0.308(7)$, whereas the lowest state $G_{2u}$ state has energy
$0.327(10)$ in the larger volume. In the smaller lattice volume,
$H_u$ and $G_{2u}$ energies are $0.310(8)$ and $0.325(9)$,
respectively. The possible spin assignments are discussed further
in the next section.

Having several good operators in each symmetry channel is
important for the success of the variational method. Based on our
present results using four $G_2$ operators, we find that the
energy of $I=\frac{1}{2}, G_{2u}$ is lighter than that of
$I=\frac{1}{2}, G_{2g}$, which is also consistent with experiment.
An earlier analysis with one $G_2$ operator found the
opposite.~\cite{Basak:2004hr}

For the $I=\frac{3}{2}, G_2$ channels, our set of
quasi-local and one-link-displaced operators does not contain a
subset that provides better signals than are obtained from a
correlation function based on a single operator. The problem is
noise. Therefore we simply fit the energies based on individual
correlation functions and pick the best-behaved one, which is
found to involve the $T_1$-displaced operator. Because no
diagonalization is involved, the energy is relatively contaminated
in early time-slices and the plateau is weak. The
$I=\frac{3}{2}, G_2$ spectrum needs further study using a larger set
of operators, such as two-link-displaced operators or operators
that have two quarks displaced from the third quark.

\section{Pattern of lowest-lying energies}
\label{sec:patterns}

 The physical spectrum of baryon excited states shows a number
 of degeneracies between states of different spins.  In
 the lattice results, this means that particular care must be
 exercised in order to identify spins because accidental
 degeneracies of two states can provide the same patterns in the
 octahedral irreps as a single higher spin state.  Given the limited energy
 resolution of our calculations, this leads to alternative
 interpretations for the spins of some excited states in the lattice results.
 It is more appropriate to compare the pattern of lattice results to the
 pattern of experimental masses and spins subduced to the irreps of
 the octahedral group.
\begin{figure}[h]
\hspace{0cm}

\vspace{.3in}
\includegraphics[width=0.5\tw]{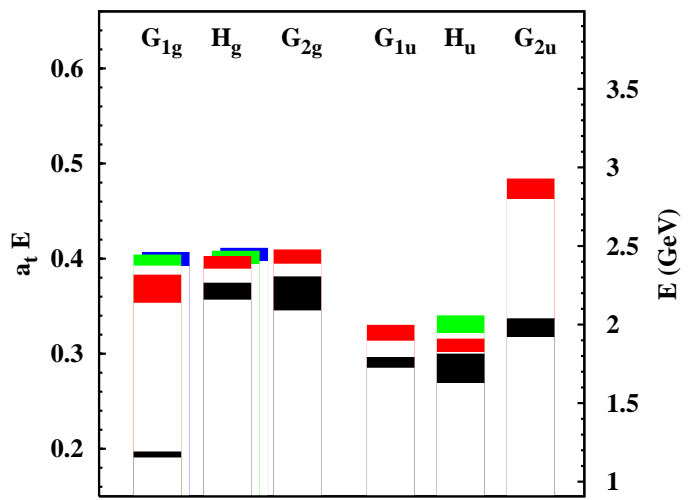}
 \vspace{0.1in} \caption{The
energies obtained for each symmetry channel of $I={1\over
2}$ baryons are shown based on the $24^3\times64$ lattice data.
The scale on the left side shows energies in lattice units and the
scale on the right side shows energies in GeV. The scale was set
using the string tension. Errors are indicated by the vertical
size of the box.} \label{fig:N24energies}
\end{figure}
Although there are substantial discretization errors with the
quark action that is used, and they could contribute differently
in the different irreps, clear patterns in the degeneracies
emerge. Focusing on the group of three positive-parity excited
states near lattice energy $a_t E = 0.36$ in
Fig.~\ref{fig:N24energies}, two interpretations are possible. A.)
The group consists of a spin $\frac{1}{2}$ state and a spin
$\frac{5}{2}$ state that accidentally are degenerate. In this case
the $G_{1g}$ state corresponds to spin $\frac{1}{2}$ and the $H_g$
and $G_{2g}$ partner states correspond to the subduction of spin
$\frac{5}{2}$. B.) The group consists of a single state with the
degenerate $G_{1g},~ H_g$ and $G_{2g}$ partner states
corresponding to the subduction of spin $\frac{7}{2}$.  Note that
these are the only possible interpretations. It is not possible
for the $H$ state to be an isolated spin $\frac{3}{2}$ state
because that would require the $G_2$ state also to be an isolated
state.  There is no interpretation for an isolated $G_2$ state.
Higher spins than $\frac{7}{2}$ would require more partner states
than are found.

In the physical spectrum of positive-parity nucleon resonances,
the lowest excited state, $N(\frac{1}{2}^+,1440)$, lies below all
negative parity states. We do not find a signal for a
positive-parity excitation that has lower energy than the
negative-parity excitations at this quark mass. The next two
excited nucleon states are essentially degenerate, namely,
$N(\frac{5}{2}^+,1680)$ and $N(\frac{1}{2}^+,1710)$, each with a
width of about 100 MeV. Spin $\frac{7}{2}$ states occur only at
significantly higher energy (1990 MeV). These states are well
separated from other states and the width of each is about 100
MeV. Primarily because of the absence of spin $\frac{7}{2}$ in the
low-lying spectrum, interpretation A.) of our lattice results is
more consistent with the pattern of physical energies and spins.

In the negative-parity spectrum, we also obtain essentially the
same results for both lattice volumes. The three lowest states
shown on the right half of Fig.~\ref{fig:N24energies} are
unambiguously identified as follows: the lowest $G_{1u}$ state
corresponds to spin $\frac{1}{2}$ and the lowest two $H_u$ states
correspond to distinct spin $\frac{3}{2}$ states. Above these is a
group of three states with roughly the same lattice energy: $a_t E
\approx 0.33$ (within errors). Again there are two possible
interpretations.  A.) The group consists of a spin $\frac{1}{2}$
state in $G_{1u}$ that is accidentally degenerate with a spin
$\frac{5}{2}$ state, the latter having degenerate partner states
in $H_u$ and $G_{2u}$.  B.) The group consists a spin
$\frac{7}{2}$ state having degenerate partner states in $G_{1u}$,
$H_u$ and $G_{2u}$.

The pattern of low-lying physical states starts with
$N(\frac{3}{2}^-,1520)$, $N(\frac{1}{2}^-,1535)$ and
$N(\frac{3}{2}^-,1700)$. These should correspond to distinct
$H_u$, $G_{1u}$ and $H_u$ states on the lattice, in agreement with
the three lowest negative-parity states in
Fig.~\ref{fig:N24energies}. The next two physical states are
$N(\frac{1}{2}^-,1680)$ and $N(\frac{5}{2}^-,1675)$, which
essentially are degenerate.  They should show up as degenerate
$G_{1u}$, $H_u$ and $G_{2u}$ states on the lattice. This pattern
of spins is consistent with interpretation A.) of the lattice
states at lattice energy $a_t E \approx 0.33$.
The pattern of energies of the physical states has
$N(\frac{5}{2}^-,1675)$ a little lower in energy than
$N(\frac{3}{2}^-,1700)$, but the lattice results at lattice
spacing $0.1$ fm place the spin $\frac{5}{2}$ state above the spin
$\frac{3}{2}$ state. Study of the continuum limit of the lattice
spectrum is required in order to resolve these issues.

\begin{figure}
\centering \ig[width=0.5\tw]{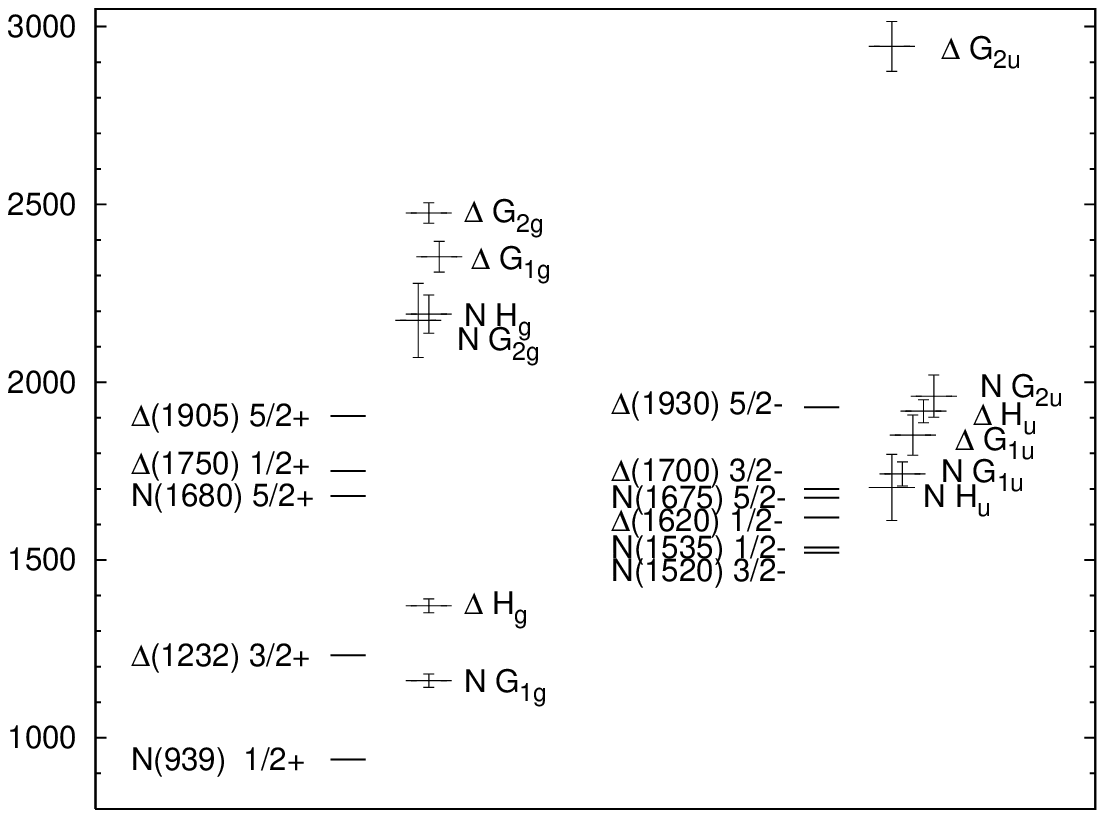} \vspace{0.1in}
\caption{The lowest energies obtained for each symmetry channel of
$I={1\over 2}$ and $I={3\over 2}$ baryons are shown in columns
2 and 4 and the experimental levels are shown in columns 1 and 3.
The $24^3\times64$ lattice data are shown and the left two columns
show the energies of positive-parity states and the right two
columns show the energies of negative parity states. The vertical
axis shows energy in MeV.} \label{fig:Phy_Lat_ND24}
\end{figure}
Figure~\ref{fig:Phy_Lat_ND24} shows the pattern of lowest-lying
energies for each irrep with the $\Delta$ states included.
Energies have been converted to physical units but the purpose of
this graph is to display the \textit{pattern} of level orderings
for low lying spin-parity channels with spins up to ${5\over 2}$,
not to make a comparison of absolute energy values with
experiment.

The spin assignments in the negative parity channels obtained in
our simulations correspond reasonably well with the physical
spectrum, \textit{i.e.}, the $N ({3\over 2}^-)$ is the lowest, the
$N ({1\over 2}^-)$ is slightly above, the $\Delta({1\over 2}^-)$ is
next, the $\Delta({3\over 2}^-)$ is next, the $N( {5\over 2}^-)$, to
which $I=\frac{1}{2}, G_{2u}$ corresponds, is slightly
above, and finally the $\Delta ({5\over 2}^-)$ is highest. One
result that does not correspond with nature is the relative order
of the $N ({5\over 2}^-)$ and the $(\Delta {3\over 2}^-)$ states,
although the error bars overlap and the energy splitting of the
mean values is only 2\%. Considering these six states, the mean
energies of $N({3\over 2}^-)$ and $N({1\over 2}^-)$ are less than the
other four. This agrees with what is predicted by the quark model
with the contact spin-spin interaction between quarks treated as a
perturbation.

In the positive-parity channels, with the exception of
the Roper resonance, our spin assignments also follow
the same orderings as occur in nature: the $N({1\over 2}^+)$ is the
lowest, the $\Delta({3\over 2}^+)$ is next, the $N({5\over 2}^+)$ is
next, for which $H_g$ and $G_{2g}$ are degenerate, the $\Delta
({1\over 2}^+)$ is next up, and finally the $\Delta({5\over 2}^+)$ is
highest. The analysis of Isgur and Karl based on the quark model
yields a nearly degenerate set of states corresponding to
$N({1\over 2}^+,1710)$, $N({3\over 2}^+,1720)$, and $N({5\over
2}^+,1680)$. Their calculation explains well the small energy
splitting, $\simeq 40\;\mbox{MeV}$, between these states. Our
results yield a fair agreement with this result. Energies of the
first excited state of $I=\frac{1}{2}, G_{1g}$, the first
excited state of $I=\frac{1}{2}, H_{g}$, and the lowest
$I=\frac{1}{2}, G_{2g}$ are $0.398(6)$, $0.395(7)$, and
$0.362(17)$, respectively. Closeness of first two energies above
is remarkable, but the third energy is significantly different.
Note that our assignments cause the lowest $N({3\over 2}^+)$
to correspond to the \textit{first excited state} of
$I=\frac{1}{2}, H_g$ because the lowest
state corresponds to the $J^P={5\over 2}^+$. Although our energy
splittings at this quark mass can be much larger than those in the
physical spectrum, if our spin assignments are correct the
ordering of energies of the lightest states in a given channel
reproduces the physical ordering.

On the other hand, comparison of the order of energy levels with
different parities provides different conclusions. For instance,
the $\Delta ({3\over 2}^-)$ energy should appear between $N
({5\over 2}^+)$ and $N ({3\over 2}^+)$ energies (here only the
lowest energies are considered), but it does not.  Further study
of the continuum limit is required in order to confirm the spin
assignments.

\section{Summary and outlook}
\label{sec:summary}

This paper presents spectra for isospin $I={1 \over 2}$ and $I={3
\over 2}$ states based on lattice QCD simulations using the
quenched approximation, anisotropic lattices with $a_s/a_t = 3$
and a pion mass of $ 490\;\mbox{MeV}$. Sets of three-quark
interpolating fields are used to construct matrices of correlation
functions and the variational method is used to extract energies.
Smeared quark and gluon fields are used to diminish the coupling
to short wavelength fluctuations of lattice QCD.  The result is
that diagonalizations are able to extract good signals for
low-lying states. We have obtained as many as 17 energies for
$I=\frac{1}{2}$ states and 10 energies for
$I=\frac{3}{2}$ baryon states, including various spin-parity
channels. The variational method determines the best linear
combinations of basis operators for energy eigenstates. For each
obtained energy eigenstate, we paid close attention to the
stability of the eigenvector components with respect to time and
with respect to the two different volumes.

Because scattering states are expected to be present in the
spectrum, we have calculated spectral weights for all states on
two volumes. However, the weights do not provide evidence for
two-particle states in our spectra. The first positive-parity
excited state in the $I=\frac{1}{2}, G_{1g}$ channel is
found to have energy significantly higher than the
experimentally-known mass of the Roper resonance. Better ways are
being developed to identify scattering states and resonances, such
as the use of operators that should have large couplings to the
$\pi-N$ states and the use of several pion masses in order to
better detect s-wave $\pi N$ states.

 Because the minimum spin that is contained in
the $G_2$ irrep is $\frac{5}{2}$, we have found strong evidence
for spin $\frac{5}{2}$ or higher for both parities in our spectra
for $I=\frac{1}{2}$. We also have found evidence for
degenerate partner states corresponding to the subduction of spin
$\frac{5}{2}$ or higher to the octahedral irreps.

The lattice results for the $I=\frac{1}{2}$ low-lying
excited states provide the correct number and pattern of
octahedral states for the subduction of the spins of the low-lying
physical states, with the exception of the Roper state. When the
$\Delta$ states are included, the overall pattern of lattice
results for the lowest energy in each channel is similar to the
pattern observed in nature.

The results shown here and in Ref.~\cite{Lichtl-Thesis} represent
a first glimpse of the pattern of nucleon and $\Delta$ excitations
as predicted by QCD, but they are based on the quenched
approximation, a 490 MeV pion mass and an action that may have
significant discretization effects. Similar calculations in full
QCD are under way with an improved action, several pion masses, several lattice
spacings and much larger sets of baryon interpolating operators.
%

\acknowledgements This work was supported by the U.S. National
Science Foundation under Award PHY-0653315, and by the U.S.
Department of Energy under contracts DE-AC05-84ER40150 and
DE-FG02-93ER-40762.

\end{document}